\documentclass[useAMS,usenatbib,twocolumn]{mn2e}
\pdfoutput=1

\usepackage{graphicx}
\usepackage{amsmath}
\usepackage{amssymb}
\usepackage{enumerate}
\usepackage{enumitem}
\usepackage{subfig}

\newcommand\ngrb{118 }
\newcommand\npul{607 }

\title{The Pulse Luminosity Function of {\em Swift} Gamma-ray Bursts}
\author[A. Amaral-Rogers et al.]
{A. Amaral-Rogers$^1$\thanks{E-mail: aar14@le.ac.uk},
R. Willingale$^1$ and P.T. O'Brien$^1$\\
$^1$Department of Physics and Astronomy, University of Leicester, LE1 7RH, UK}

\begin{document}
\date{Accepted 2016 September 19. Received 2016 August 28; in original form 2016 March 7}

\pagerange{\pageref{firstpage}--\pageref{lastpage}} \pubyear{2016}

\maketitle
\label{firstpage}

\begin{abstract}

The complete Swift BAT and XRT light curves of 118 GRBs with known redshifts
were fitted using the physical model of GRB pulses by Willingale et al. (2010)
to produce a total of 607 pulses. We compute the pulse luminosity function
utilising three GRB formation rate models: a progenitor that traces the cosmic
star formation rate density (CSFRD) with either a single population of
GRBs, coupled to various evolutionary parameters, or a bimodal population
of high and low luminosity GRBs; and a direct fit to the GRB formation
rate excluding any a-priori assumptions. 

We find that a single population of GRB pulses with an evolving luminosity
function is preferred over all other univariate evolving GRB models, or
bimodal luminosity functions in reproducing the observed GRB pulse L-z
distribution and that the magnitude of the evolution in brightness is
consistent with studies that utilise only the brightest GRB pulses. We
determine that the appearance of a GRB formation rate density evolution
component is an artifact of poor parameterisation of the CSFRD at high
redshifts rather than indicating evolution in the formation rate of early
epoch GRBs. We conclude that the single brightest region of a GRB lightcurve
holds no special property; by incorporating pulse data from the totality of
GRB emission we boost the GRB population statistics by a factor of ~5, rule
out some models utilised to explain deficiencies in GRB formation rate
modelling, and constrain more tightly some of the observed parameters of GRB
behaviour.
 
\end{abstract}

\section{Introduction}
\label{s1}
The luminosity function (LF) is a powerful tool for population analysis and,
when applied to Gamma-ray bursts (GRBs), is used to verify theoretical models
of the physical processes that go into forming GRBs; and as a benchmark for 
observation rates of future GRB missions, and gravitational wave detection
likelihoods. The luminosity function does however require a precise measure of
the distance to the GRB in order to convert from the observed flux to the
rest-frame luminosity. In the era of {\em Swift} observations, over one
thousand GRBs have been observed with approximately 1/3 having an associated
redshift, meaning that GRB luminosity functions are currently built upon
relatively small sample sizes compared to other luminosity functions found in
astrophysics. The standard technique for generating a luminosity function for
GRBs is to utilise either a time-averaged luminosity, or the brightest part of
a burst, as the characteristic luminosity of the GRB; and in such cases where
there is little, or no, variation in the lightcurve such an approach is
acceptable. The majority of {\em Swift} GRBs, however, show significant
variation in their lightcurves with multiple peaks in the early prompt and
late-time emission that, in some cases, are of comparable brightness to the
most luminous part of the GRB lightcurve. In this paper we follow a different
approach. Using a physically motivated model for the prompt and high-latitude
emission from GRBs \citep{2009MNRAS.399.1328G, 2010MNRAS.403.1296W} we fit the
lightcurves of \ngrb long Gamma-ray bursts (LGRBs) to produce a total of \npul
individual Gamma-ray pulses and X-ray flares. The wealth of information stored
within these other, less-luminous, pulses are utilised to produce a GRB pulse
luminosity function; of which the more conventional LGRB LF can be considered
as a high-luminosity subset. As with papers investigating the GRB luminosity
function, we find discrepencies between observed, and theoretical LF models
that require additional evolutionary effects to correct for. In this paper we
evaluate extensive solutions to these discrepencies: additional luminosity,
rate, and metallicity density evolution; bimodal low/high luminosity
functions; and direct fitting of GRB formation rates. We find that in all
cases a model that incorporates evolution in the break luminosity, such that
higher redshift GRBs are more luminous, is preferred over one that does
not. We also do not see strong evidence of a divergence between the cosmic
star formation rate density and the GRB formation rate, nor any compelling
evidence of a separate population of high and low luminosity GRBs. Our
findings are broadly consistent with GRB LF studies whilst producing better
defined evolutionary parameters, suggesting that there is nothing special
about the single brightest pulse and that studies into GRB population
behaviour should include all the pulse information available.    

Prior to the launch of {\em Swift} \citep{2004ApJ...611.1005G}, the number of
LGRBs localised to a suitably fine error circle on the sky such that follow-up
observation could find an associated host galaxy or afterglow was small; out
of some 2704 GRBs detected by BATSE \citep{2013ApJS..208...21G} only a handful
had a measured redshift, made possible only due to simultaneous detection of
the burst by other Gamma-ray missions with greater localisation
abilities. Given the lack of real redshifts, many authors instead sought 
to derive pseudo-redshifts using properties of the LGRB lightcurves in order
to derive a LGRB luminosity function; the most popular of which
included the lag-luminosity relationship \citep{2000ApJ...534..248N,
2006ApJ...642..371K}, the variability-luminosity
relationship \citep{2000astro.ph..4176F, 2001ApJ...552...57R,
2002ApJ...574..554L, 2003MNRAS.345..743W}; and the Amati
relationship \citep{2002A&A...390...81A, 2003A&A...407L...1A,
2004ApJ...611.1033F, 2004ApJ...609..935Y, 2007ApJ...656L..49S, 
2009MNRAS.396..299S}. The large intrinsic scatter within these relationships
produces however a redshift distribution that, whilst arguably it represents
that of the true LGRB redshift distribution, also shows significant
uncertainty in the fitted parameters. With the launch of the {\em Swift}
mission, with its fast slew rate and accurate on-sky localisation, suddenly a
large proportion of GRBs being detected had associated photometric and/or
spectroscopic redshifts of either the host galaxy or the GRB's X-ray afterglow
within a day or two of initial observation. Early {\em Swift} GRB LF papers
continued to develop the LGRB luminosity function by utilising either small
numbers of LGRBs with measured redshifts \citep{2008MNRAS.388.1487L,
  2008ApJ...673L.119K}, with poor constraints on fitted parameters; or by
artificially boosting the LGRB redshift sample by combining real and pseudo
redshift data from {\em Swift} and BATSE \citep{2010ApJ...711..495B,
  2012ApJ...749...68S}. Over time, the sample size of GRBs with observed
redshifts has increased\footnote{As of January 2016 over 1000 GRBs have been
  observed by {\em Swift} with 295 GRBs having associated redshifts.} and
contemporary GRB LF studies utilise larger datasets with entirely observed
redshifts \citep{2010MNRAS.406.1944W, 2011MNRAS.416.2174C,
  2012ApJ...744...95R, 2013MNRAS.428..167H, 2014arXiv1412.3969G,
  2015arXiv150401414P, 2015arXiv150605463P, 2015arXiv150401812Y,
  2016arXiv160107645D}. 

Throughout these earlier studies the emphasis has been
on trying to extract information about the average behaviour of the LGRB as a
whole; in general, characterising the luminosity of a GRB using the flux of
the single brightest peak in the lightcurve binned in 1 second bins (see for
example \cite{2004ApJ...609..935Y}). In reality the lightcurve of a GRB is
complex and the scale of variability between lightcurves is bewildering. The
luminosity and total energy output of GRBs spans many orders of
magnitude. Whilst some bursts consist of a single Fast Rise Exponential Decay
(FRED) profile, others have multiple peaks; some are very spikey with rapid
variations while others have a much smoother profile. Many lightcurves display
astonishing chaotic time-variability, continually varying between bright,
short peaks and low troughs where in some cases the flux drops below the
detection threshold for a while before flaring up again. The current paradigm
is of the totality of the prompt emission constructed of simple pulses 
\citep{2005ApJ...627..324N,2010MNRAS.403.1296W}. Lasting from fractions of a
second, to minutes in duration, these pulses are independent of each other but
with many overlapping to some degree to produce the incredible lightcurve
variation observed. Late-time X-ray flares \citep{2007ApJ...671.1903C,
  2007ApJ...671.1921F} seen above the afterglow emission hundreds, and in some
cases thousands of seconds, after the initial trigger appear to follow the
same mechanism as the prompt emission pulses and can be considered as the
lower energy tail of a unimodal pulse energy distribution\footnote{Many
  early-time X-ray flares have an observable Gamma-ray counterpart in the BAT
  however the majority of late-time X-ray flares would have Gamma-ray fluxes
  well below the BAT detection limit.}. Given the wealth of information
contained within GRB lightcurves, the approach taken by the authors is to fit
each individual pulse using a physical model characterised by a few simple
parameters: the peak flux, a characteristic time scale, a rise time as a
fraction of the characteristic time scale, and a spectral shape; the model
furthermore incorporates spectral evolution such that the rise and decay to
and from the peak of each pulse depend on the changing X-ray spectrum. Instead
of the single data point extracted by more conventional GRB luminosity
function studies, we fit, on average, 5 pulses per GRB, with the more variable
lightcurves containing pulses numbering in the tens, significantly increasing
our sample size over single pulse studies. Using the measured redshift, the
peak flux, and spectrum we derive the rest-frame bolometric luminosity for
each pulse, and use the totality of our data to construct and evaluate various
GRB pulse luminosity functions.  

The structure of this paper is therefore as follows: we discuss the selection
criteria for our GRB sample; the physical model for pulses, and flares; and
the fitting technique in Section 2. We outline the various methods for
constructing a luminosity function in Section 3; and in Section 4 we discuss
the Markov Chain Monte Carlo (MCMC) routine utilised to fit our luminosity
function model parameters. Sections 5, 6, and 7 are discussions on the results
from models convolved to the cosmic star formation rate density with either: a single population of GRB progenitors (Type I models), or two separate populations of high, and low luminosity GRBs (Type II models); and GRB formation rate models that rely on no prior assumptions on progenitor mechanisms (type III models). Section 8 compares the LGRB formation rates with
the observed cosmic star formation rate density and we finally conclude our
findings in Section 9.

\section{Lightcurve fitting}
\label{s2}
We analysed the lightcurves of LGRBs with associated redshifts observed
by both the {\em Swift} BAT \citep{2005SSRv..120..143B} and
XRT \citep{2005SSRv..120..165B}, during the period bordered by LGRBs 050126
and 110503A inclusively, using the pulse, and afterglow model procedure
described in \citet{2010MNRAS.403.1296W}; a description 
of which is summarised in Section 2.1. Our definition as to which class, long
or short, a GRB belongs to is based solely on the $T_{90}>2s$ descriminator
found by \cite{1993ApJ...413L.101K} and as such may contain some bursts which
exhibit the spectral characteristics of short GRBs whilst residing solidly in
the LGRB duration regime: so called "extended emission" (EE) bursts
(e.g. \citet{2006ApJ...643..266N}, \citet{2014MNRAS.438..240G}). EE GRBs are a
small, and contentious subsample of our GRB dataset and may contribute a
small bias in our LGRB formation rates due to the propensity of SGRBs to be
detected at low redshifts \citep{2014MNRAS.442.2342D}; this is however
mitigated to a large extent by the overall number of fitted pulses and as such
is not of significant concern.     
    
\noindent
The following criteria were used to determine the suitability of each GRB for
pulse fitting, with the bracketed figures denoting the number of GRBs rejected by the criteria:
\begin{description}
\item [-] the GRB has an observed redshift and a $T_{90}>2s$ (13 GRBs rejected);  
\item [-] sufficient statistics in the BAT lightcurve to define at least 1 pulse
profile (29 GRBs rejected); 
\item [-] a BAT lightcurve in which pulses are reasonably well defined (3 GRBs rejected);
\item [-] early data from the XRT so that the decay of the pulses is well
constrained (22 GRBs rejected); and
\item [-] XRT data which provide good definition
of X-ray flares, avoiding flares for which profiles are incomplete or
broken by orbit gaps etc.. (22 GRBs rejected)
\end{description}

We evaluate any redshift bias that the various rejection criteria may accidentally introduce into our LGRB dataset by computing the 2-sample Anderson-Darling (AD) statistic \citep{And_Dar_1952, 10.2307/2281537, darling1957} \footnote{The Anderson-Darling test statistic is a modified Kolmogorov-Smirnov test statistic, and is preferred due to its greater sensitivity to differences in the tails of distributions, and its ability to sence differences between very large datasets.} on the redshift distributions of accepted GRBs and of GRBs which failed the rejection criteria, under the null hypothesis that both are drawn from the same population. As we are calculating the likelihood of two distributions being drawn from the same parent distribution, no assumptions are required for the shape of the parent; this is not the case were we calculating a one-sample test. We utilise the k-sample Anderson-Darling test codified in the SciPy stats package \citep{SciPy_Cite}, which is based on the work by \citet{scholz1987k}. The critical significance values are modelled as a third order polynomial, and interpolated over a percentile grid of [0.75, 0.90, 0.95, 0.975, 0.99]; outside of this range, the $P$-values are extrapolated and, as such, come with large uncertainties the further away one gets. We therefore quote the calculated AD statistic and the appropriate significance level to which the null hypothesis may be rejected.

We can reject the null hypothesis that the redshift distribution of GRBs with a $T_{90}<2$ s is drawn from the same distribution as LGRBs, as the AD statistic of 11.377 corresponds to a significance level of $P>0.99$; this result is entirely expected given our understanding of SGRBs and LGRBs. We also find that we can accept the null hypothesis for both the rejection criteria of early XRT data, and complete XRT flares, having been drawn from the same parent population as our sample of LGRBs: the AD stats are 0.825, and -0.875 respectively, which corresponds to $P$-values of $P=0.85$ and $P\ll 0.75$ \footnote{The threshold for rejection is often given at the arbitrary $P$-value of $>0.95$.}. The criteria for a minimum of one pulse in the BAT, and a well defined BAT pulse are, like the $T_{90}$ criterion, both rejected with $P$-values of $>0.99$. It is highly likely, however, that short GRBs have a poorly observed BAT regime, as the hardness of short GRBs does not lend well to detection by the relatively soft BAT passbands. Calculating a 3-sample AD statistic shows that the null hypothesis that GRBs which fail these three criteria are drawn from the same population can be accepted, with an AD statistic of -0.079 corresponding to a $P$-value of $\ll 0.75$; this suggests that these rejected GRBs are indeed most likely short, and by excluding them we do no insert any significant bias to our LGRB dataset.  

In summary, out of 187 LGRBs with associated redshifts covering the 76 month period from GRB 050126 to GRB 110503A with $T_{90}>2s$, \ngrb GRB lightcurves were deemed suitable and fitted with \npul pulses: a completeness of $\sim63\%$. As a comparison study, \citet{2012ApJ...749...68S} utilised GRBs spanning an almost identical time period as our own and, after applying their selection criteria, drew a population of 58 LGRBs of which 52 have measured redshifts: a completeness of $\sim39\%$ \footnote{The \citet {2012ApJ...749...68S} completeness is derived from the 132 available redshifts that were available at the time of that paper's writing. Whilst the majority of GRB redshifts are released within a few days of initial observation, some are derived, or updated, only after extended follow up observations months, or years, after the initial burst; in all cases we endeavour to obtain the most up to date redshift information available.}. 

\subsection{The Pulse Model}

A photon, emitted from the source when a shell is ejected from it, arrives
in the observer frame at time $T_{ej}$ which can be thought of as the observed
ejection time of the shell. The initial radial time, $T_{0}$ is assumed such
that the time at which the first photons emitted from the emission region at
radius $R=R_{0}$ reaches the observer at $T=T_{ej}+T_{0}$. Likewise the final
radial time, $T_{f}$, emitted from the emission region at $R=R_{f}$ reaches
the observer at time $T=T_{ej}+T_{f}$. For compactness of equations we also define two normalised times:

\begin{equation}
\begin{array}{ll}
\bar{T}\equiv \frac{T-T_{ej}}{T_{f}}; & \bar{T_{f}}\equiv \frac{T_{0}}{T_{f}}.
\end{array}
\label{eq1}
\end{equation}

\begin{figure}
\begin{center}
\includegraphics[width=\linewidth, bb = 0 0 450 450]{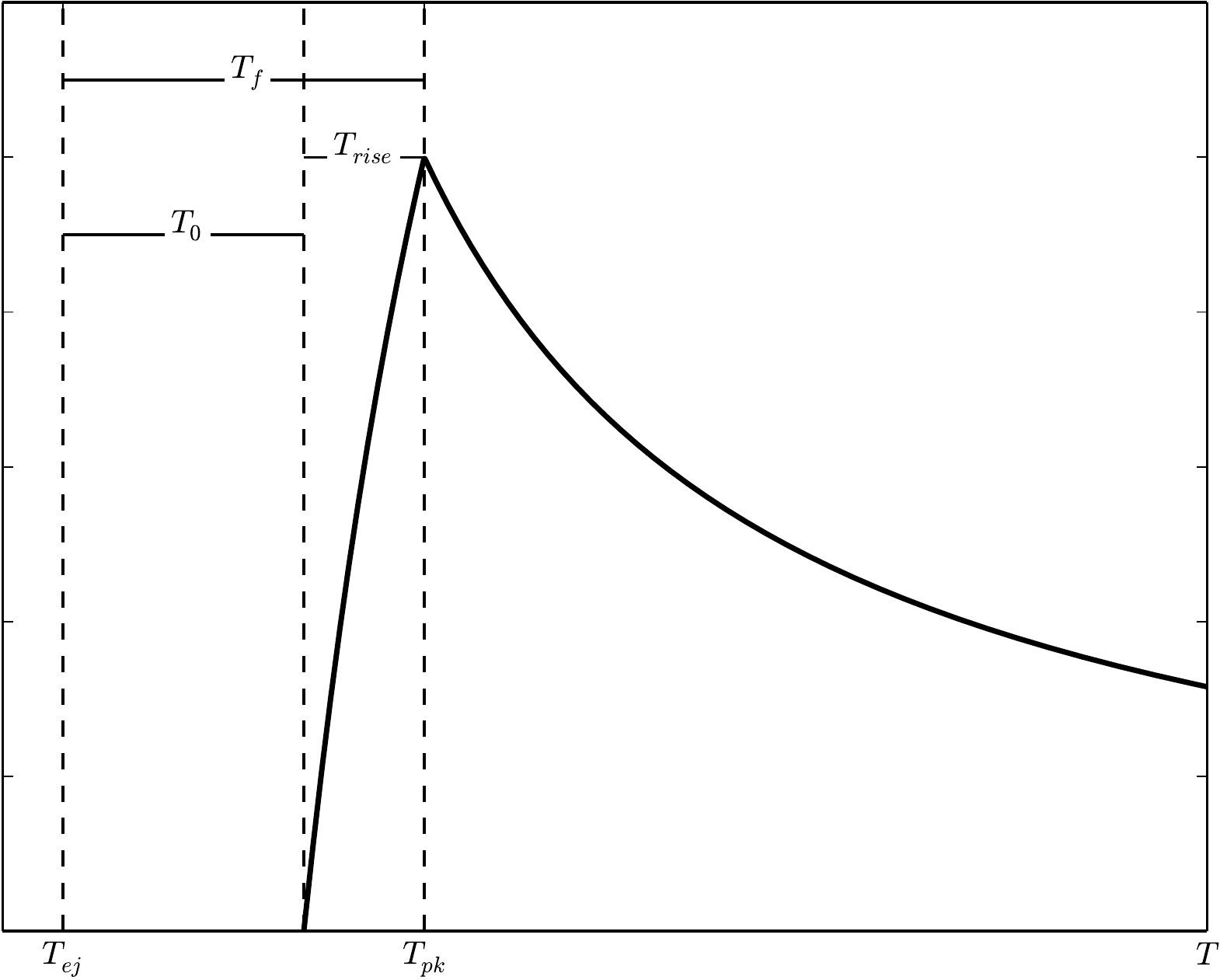}
\end{center}
\caption{A schematic for the pulse profile, $P(\bar{T},\bar{T_{f}})$, showing
the characteristic timescales $T_{f}$, $T_{0}$, and $T_{rise}$; and the times
$T_{ej}$ and $T_{pk}$.}    
\label{fig1} 
\end{figure}

Integrating the comoving luminosity over the equal arrival time surface (EATS) in combination with the model spectrum, $B(q)$ (see Equation \ref{eq4}), allows us to derive the flux in terms of number of photons, $N$ per unit energy, $E$, area $A$ and time $T$:

\begin{equation}
\frac{dN}{dEdAdT}(E, T\geq T_{0}+T_{ej})=P(\bar{T},\bar{T_{f}})B(q);
\label{eq2}
\end{equation}

\noindent
the pulse profile, $P(\bar{T},\bar{T_{f}})$, is given by: 

\begin{multline}
P(\bar{T},\bar{T_{f}}) = P_{norm}\bar{T}^{-1}\left[\left(min(\bar{T},1)^{a+2} -
  \bar{T_{f}}^{a+2} \right) \right]
\label{eq3}
\end{multline}

\noindent
where the pulse profile includes a normalisation parameter,
$P_{norm}=\left(1-\bar{T_{f}}^{a+2} \right)^{-1}$ such that the value of
$P(\bar{T},\bar{T_{f}})$ is $1$ at $T=T_{ej}+T_{f}$. The section in square
brackets in Equation \ref{eq3} models the rise in the pulse, in this case
controlled by the  temporal index $a$ and the timescales $\bar{T}$ and
$\bar{T_{f}}$. A schematic of the pulse is shown in Figure \ref{fig1} with
characteristic timescales denoted.

Whilst the temporal characteritics of GRBs display remarkable variation, their spectral profiles are far less varied. The spectra of GRB pulses during the prompt phase are distinctly non-thermal; as such, we model each individual pulse by a Band function, given by Equation \ref{eq4}:

\begin{equation}
B(q)=B_{n}\left\{
\begin{array}{ll}
q^{b_{1}-1} e^{-q} & q \le b_{1}-b_{2}  \\
q^{b_{2}-1}(b_{1}-b_{2})^{b_{1}-b_{2}} e^{-(b_{1}-b_{2})} & q > b_{1}-b_{2}  \\
\end{array}
\right.
\label{eq4}
\end{equation}

\noindent
where $b_{1}$ and $b_{2}$ are the low and high energy spectral indexes
respectively, $B_{n}$ is the normalisation parameter, and $q=(E/E_{f})\bar{T}=E/E_{c}$. The cutoff energy at time $T_{peak}=T_{ej}+T_{f}$, denoted as $E_{f}$, coincides with the maximum emission from the pulse where the spectral profile is best constrained. The hard-to-soft evolution seen in GRB pulses are modelled through an evolving characteristic energy, $E_{c}$, such that $E_{c}(t) = E_{c}(T - T_{ej}) = E_{f}(\bar{T})^{-1}$, where the strength of the spectral evolution is a result of assuming synchrotron dominated emission in the fast-cooling regime (see Willingale et al. 2010 for derivations). Whilst there is, in some rare cases, evidence of an underlying and statistically significant thermal component within the spectra of a few GRBs (see \citealp{2011ApJ...727L..33G} for example), we feel that incorporating an additional thermal component would be an exercise in diminishing returns.  

\subsection{Fitted parameters}

The time of the peak with respect to the trigger, $T_{pk}$, was initially
set by eye and then allowed to float to find the best value. In the final fits
$T_{pk}$ was then fixed at the best value found whilst in all cases the
characteristic time $T_{f}$ was allowed to float. Instead of fitting the rise
time directly it was expressed as a fraction of $T_{f}$,
$T_{rise}=f_{r}T_{f}$, and it is this fraction which was fitted with the
constraint of $0<f_{r}<1$. Using this fraction provides a simple way to avoid
unphysical rise time values not allowed by the model. In many bursts
significant emission is seen before the trigger and in some bursts the first
peak may have a large negative $T_{pk}$ value. On completion of the final fit
we redefined the zero time as the start of the emission in the model given by
the start of the rise of the first pulse, $T_{zero}=T_{pk}(1)-T_{rise}(1)$. We
then offset the peak times of all the pulses in the burst to
$T_{peak}=T_{pk}-T_{zero}$ so this represents the time since the start of the
burst for each pulse. 

For all pulses the low-energy index of the Band function was fit whilst the difference between the low and high spectral indicies was fixed at $b-d=10$. This effectively reduces the Band function to a simpler power law with an exponential cut-off. For a few pulses the count rate in the higher energy channels was effectively zero and the spectral index was very poorly determined. In such cases the lower spectral index was therefore constrained to $b> -2.5$. Because of the relatively soft energy bandwidth of the {\em Swift} BAT, and the signal to noise of the measured light-curves, a powerlaw with exponential cutoff produces comparable quality of fits to Band functions without being so computationally demanding. 

For the majority of GRB pulses the cut-off energy of the Band function lies outside the passband of the BAT; in such cases we fix the cutoff energy of the Band function at $T_{peak}$ at $E_{fz}=500$ keV in the source frame of the burst, corresponding to $E_{f}=500/(z+1)$ keV in the observer frame, similar to the fixed cutoff energies utilised by other studies, e.g. \citealt{2004ApJ...611.1033F, 2005MNRAS.364L...8N}. For some pulses however, with good statistics and energy coverage (including both the BAT and XRT data), it was possible to constrain $E_{f}$ by the fitting to some other value (usually a lower energy). As joint analysis of the spectra of GRB pulses observed simultaneously by {\em Swift} and other satellites such as Fermi, Suzaku, and Konus-Wind are rare and are often based on a few GRBs (see for example \citealt{2009ApJ...704.1405K}), we cannot directly compare spectral fits on a pulse-by-pulse basis for the majority of our \npul pulses. We instead compare the spectral characteristics of the prompt emission pulses utilised within this paper with the time-averaged spectral parameters observed by other space-based gamma-ray, and X-ray observatories with wider energy passbands than the {\em Swift} BAT; out of \ngrb GRBs, 51 were observed by other missions, totalling 183 prompt-phase pulses. 

Although not strictly equivalent, as the totality of the GRB prompt emission is a convolution of many constituent pulses, such a comparison can reveal any significant differences. To this end we define a deviation metric for parameter $P$ such that $\Delta P = |P_{Swift}-P_{other}|/ \sigma_{combined}$ where $\sigma_{combined}$ is the resulting uncertainty of the two measurements combined in quadrature ($\sigma_{combined}^{2} = \sigma_{Swift}^{2}+\sigma_{other}^{2}$), and a $\Delta P < 1$ denotes a parameter that is within the combined $1\sigma$ uncertainties. We find good agreement between our pulse spectral parameters and those of the time-averaged GRB spectra, with the median deviation in the spectral indexes, and peak energies of $\Delta B_{1}=1.25^{+1.89}_{-0.70}$, and $\Delta E_{peak}=0.70^{+0.52}_{-0.35}$; where the subscripts/superscripts denote the $25^{th}$ and $75^{th}$ percentiles respectively. Such differences in the spectral parameters of our pulses and the time-averaged GRB prompt emission will produce K-correction factors which may vary significantly, and by extention, produce bolometric rest-frame luminosities that are widely different. We therefore calculate and compare the K-corrections one would derive assuming a power-law with exponential cutoff spectrum for both measurements. We observe a median deviation between the two broadband observations on the scale of $\Delta K_{corr} = 0.28^{+0.24}_{-0.15}$; we conclude therefore that the effect of introducing a fixed cutoff energy in the spectra of our pulses is negligible. 

\begin{figure}
\centering
\subfloat{\includegraphics[width=0.9\linewidth, bb = 0 0 600 800]{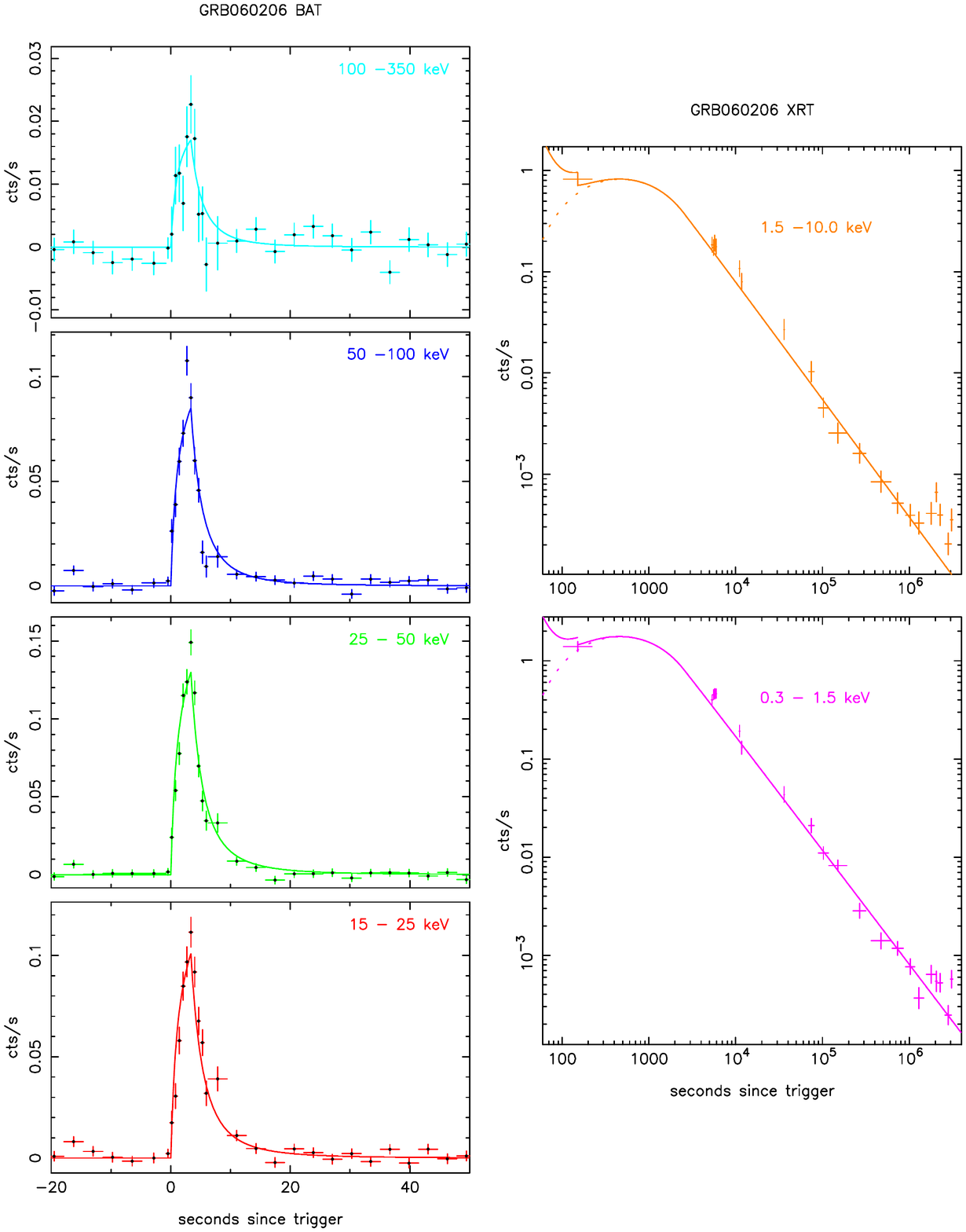}}\\
\subfloat{\includegraphics[width=0.9\linewidth, bb = 0 0 600 800]{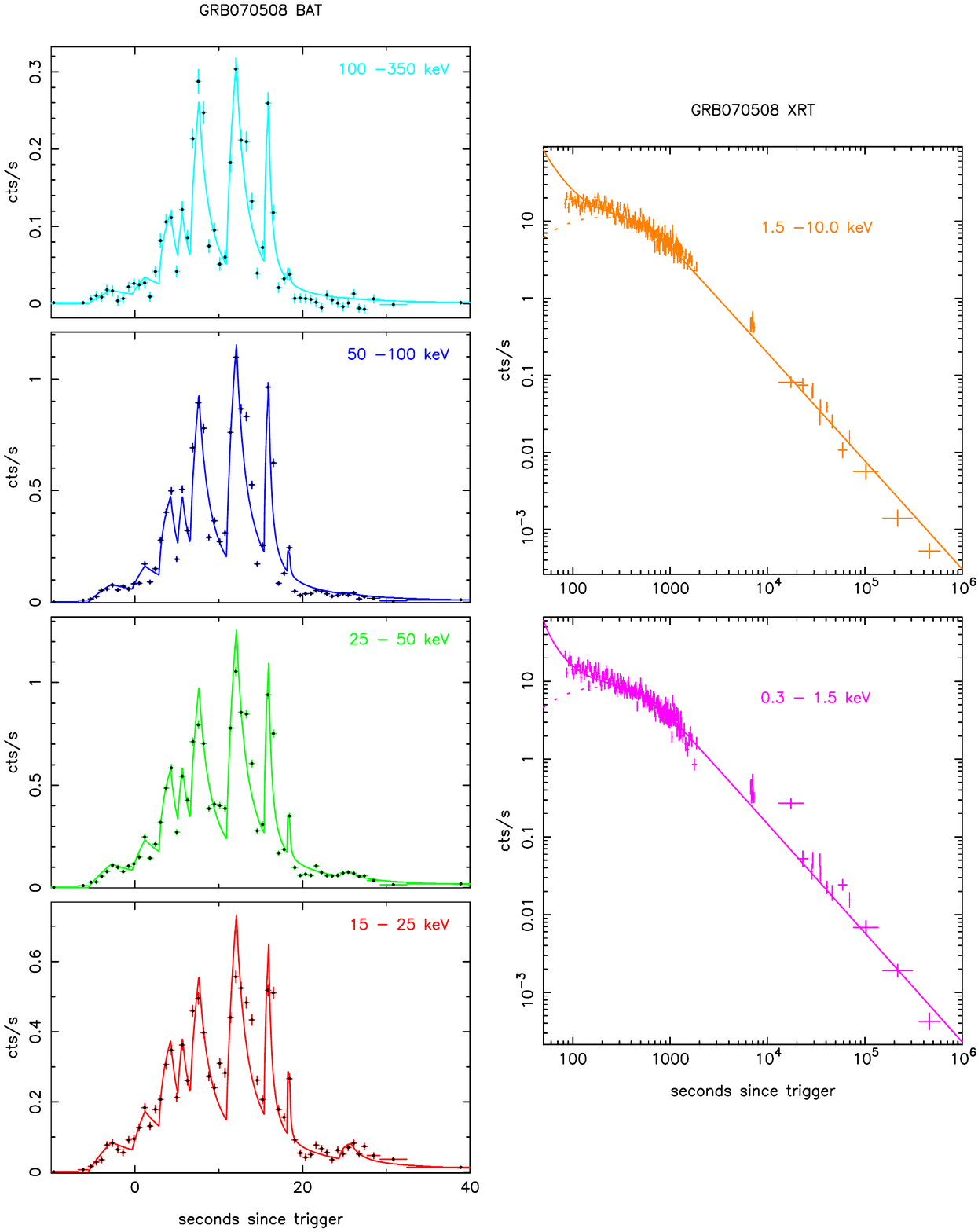}}
\caption{The pulse model fits for GRB 060206 (top) and GRB 070508 (bottom). The left columns corresponds to the BAT passbands of 100 - 350 keV (cyan), 50 - 100 keV (blue), 25 - 50 keV (green), and 15 - 25 keV (red); the right columns are the XRT passbands of 1.5 - 10 keV (orange), and 0.3 - 1.5 keV (magenta).}    
\label{fig2} 
\end{figure}

In general the pulse profiles are well matched by the model but Figure \ref{fig2} illustrates
typical deficiencies in the fitting (see \citealp{2010MNRAS.403.1296W} for
further discussion on the various fitting pitfalls). For GRB 060206 the pulse
decays more rapidly in the hard band than predicted and there are minor
excursions away from the model over several of the pulses in GRB 070508. For
many bursts there are a couple of data points in the harder bands which form a
spike which is not fitted by the model profile. A few points in the decay of
the afterglow in GRB 070508 are well above the model in the softest band and
the very late points of the afterglow in GRB 060206 are poorly fit. In these
fits such errors can't be accomodated for by the addition of more pulses and
subsequently contribute to some of the larger $\chi^{2}$ values
obtained. Despite these issues we tried to fit every pulse-like feature in all
the lightcurves and the combination of the pulse model plus afterglow
accounts, in most cases, for all the emission detected from all the
bursts. \footnote{The data used throughout this paper will be available in an
upcoming GRB components catalogue paper by the authors.}

\section{Modelling the GRB Luminosity Function}
\label{s3}
We note that the nomenclature of "luminosity function" in
reference to GRBs refers specifically to the GRB luminosity probability
density function (PDF); to obtain what is in general analogous to the LFs
found in other areas of astrophysics one must convolve the GRB luminosity PDF
with the cosmic GRB formation rates. Any subsequent reference to the GRB
luminosity function in this paper will follow this convention and refer to the
GRB luminosity PDF. Throughout this paper we used the formulation of comoving
distance, volume and luminosity distance given by \cite{1999astro.ph..5116H}
utilising the seven-year WMAP cosmological parameters of $H_{0}=71$ km
s$^{-1}$ Mpc$^{-1}$, $\Omega_{m}=0.27$, $\Omega_{k}=0$ and $\Omega_{\Lambda}=0.73$
\citep{2011ApJS..192...16L}. All errors quoted in this paper are to the
$1\sigma$ confidence interval in line with the majority of GRB LF literature.

Throughout this paper we discuss reproducing the GRB pulse luminosity function though a variety of models which, in some cases, include various sub-models. Type I models invoke a cosmic star-formation rate coupled to a single population of GRB progenitors (Section 3.2); type II models are similar to type I save for the separation of GRB progenitors into low, and high-luminosity populations (Section 3.1.1); whilst type III models are direct fits to GRB formation rates and exclude a-priori assumptions about the nature of GRB progenitors (Section 3.2). Models I, and III are further explored through the inclusion of various extra evolutionary effects (see Section 3.6) and are summarised as:

\begin{description}[style=unboxed]
\item[Type I-1:] no evolution in either the break of the pulse LF, $L_{break}$ ($\delta=0$),
  or the GRB formation rate, $K_{GRB}$ ($\gamma=0$ or $Z/Z_{\odot}=\infty)$; 
\item[Type I-2:] evolution of only the GRB formation rate, $K_{GRB}$ ($\gamma\neq0$);
\item[Type I-3:] evolution of only the break, or cutoff, of the luminosity function,
  $L_{break}$ ($\delta\neq0$);
\item[Type I-4:] evolution of the GRB formation rate through the presence of metallicity
  density evolution ($Z/Z_{\odot}\neq\infty$);
\item[Type I-5:] both $L_{break}$ and $K_{GRB}$ are free to evolve ($\delta$ $\&$
  $\gamma\neq0$). 
\item[Type III-1:] no evolution in the break of the pulse LF, $L_{break}$ ($\delta=0$);
\item[Type III-2:] evolution in the break of the pulse LF, $L_{break}$ ($\delta\neq0$);
\end{description}

\begin{figure}\begin{center}
\includegraphics[width=\linewidth, bb = 0 0 650 500]{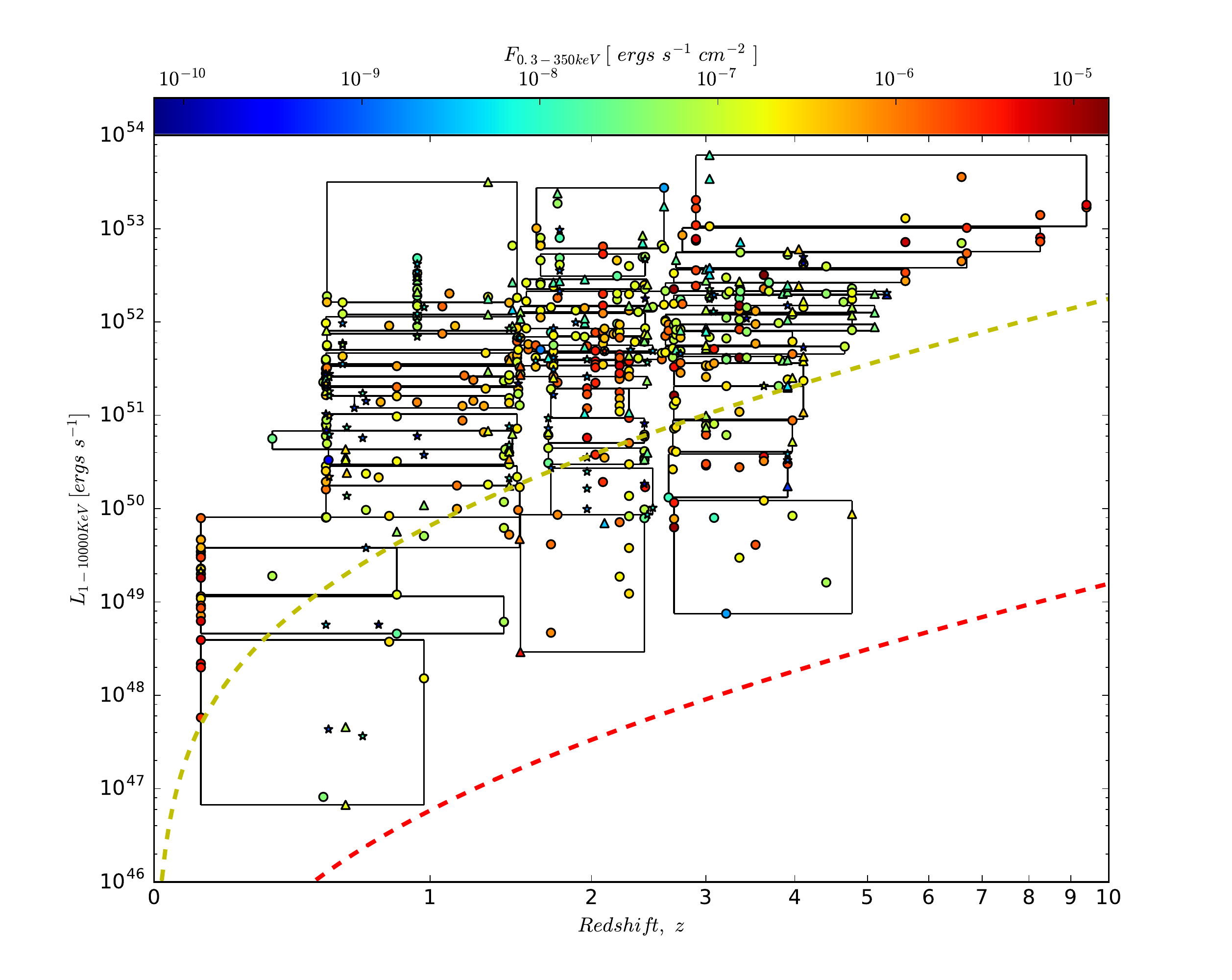}
\end{center}
\caption{The distribution of peak luminosities as a function of
    redshift, $N(L,z)$. The circles and stars indicate pulses detected
    only with the BAT or XRT instruments respectively with triangles denoting pulses detected by both instruments simultaneously. BAT observed
    pulses are visible below the BAT detection limit as fainter pulses
    co-added to prior, brighter pulses can be "boosted" into the detection
    range. Pulse colours denote the integrated peak flux in the observed 0.3 - 350 keV passband. The overlain boxes show the corresponding $J_i$ sets used to
    determine each bin's limits with an explanation to the derivation of set
    limits outlined in Section 4.1. The yellow dashed and red dashed lines
    show the detection threshold of the bolometric luminosity as a function of
    redshift, with a burst detection threshold in the BAT (15-350 keV) of 0.1
    photons cm$^{-2}$ s$^{-1}$, equivalent to $8\times 10^{-9}$ ergs cm$^{-2}$
    s$^{-1}$ \citep{2008ApJS..175..179S} and an approximate XRT (0.3-10 keV)
    pulse/flare detection limit of $3\times 10^{-12}$ ergs cm$^{-2}$
    s$^{-1}$. In practice the XRT detection limit depends on the brightness of
    the afterglow component and the time since trigger so the lower detection
    threshold varies considerably from burst to burst. Few pulses are seen
    near the XRT threshold and the detection likelihood of pulses is discussed
    in Section 3.5. The K-correction to convert the BAT \& XRT flux detection
    limits to that of bolometric luminosities utilises the average spectral
    parameters of the Band function derived from our dataset of \npul pulses
    with $b=-1.57$, $d=-11.60$ and $E_{f}=183$ keV.} 
\label{fig3}
\end{figure}

The observed distribution of pulse bolometric luminosities, $N(L,z)$, by
definition spanning the energy band of 1 - 10000 KeV\footnote{The bolometric
luminosity of each individual pulse is derived from applying a K-correction to
the pulse flux using the spectrum at peak time as a fiducial spectrum.}, is
displayed in Figure \ref{fig3}. Pulses for which the peak only
appears in the BAT or XRT lightcurves are shown as circles and stars
respectively, whilst pulses observed simulatenously by both instruments are
denoted by triangles. The $N(L,z)$ distribution displays a wide range of
brightnesses for prompt emission pulses, and late time X-ray flares; and
whilst the very brightest of pulses ($L_{1-10000KeV} > 1\times10^{53}$ ergs
$s^{-1}$) are exclusively from the prompt emission, the X-ray flares and
prompt emission pulse luminosity distributions are indistinguishable from each
other.

The standard procedure for relating the observed distribution of LGRBs to the
comoving burst formation rate (see for example \cite{2000astro.ph..4176F,
2002ApJ...574..554L, 2007ApJ...656L..49S, 2010ApJ...711..495B,
2012ApJ...749...68S}) is given by:

\begin{equation}
N(L,z) dzdL
  = \phi(L,z) \left[D(L,z) \frac{\Delta\Omega\Delta T\Psi_{*}(z)}{(z+1)} 
  \frac{dV_{c}(z)}{dz} \right]dzdL
\label{eq5}
\end{equation}

\noindent
where the observed distribution of LGRB bursts, $N(L, z)$, is a convolution of
the comoving burst formation rate, $\Psi_{*}(z)$, the comoving volume element,
$dV_{c}(z)/dz$, a detection probability profile, $D(L,z)$, and the GRB
luminosity probability density function, $\phi(L,z)$. The factor of
$1/(z+1)$ corrects for cosmological time dilation whilst $\Delta\Omega$ and
$\Delta T$ are the terms correcting for the field of view of the BAT and the
total duration our GRB sample covers. 

\subsection{Luminosity Function}

The functional forms for LGRB LFs represented in the {\em Swift} literature are
predominantly that of a broken power-law (sometimes with a smoothed
transition between low and high luminosity
regions) \citep{2002ApJ...574..554L, 2007ApJ...662.1111L,
2010ApJ...711..495B, 2011MNRAS.416.2174C, 2012ApJ...749...68S}, or a power-law
with an exponential cutoff \citep{2007ApJ...656L..49S, 2009MNRAS.396..299S, 2011MNRAS.416.2174C, 2012ApJ...749...68S}. In this paper, to
ensure completeness, we utilise both a broken power-law (BPL), 

\begin{equation}
\phi(L,z) =  L_{n}\left\{
\begin{array}{ll}
\left(\frac{L}{L_{break}}\right)^{\alpha} & L \leq L_{break} \\
\left(\frac{L}{L_{break}}\right)^{\beta}  & L > L_{break}  \\
\end{array}\right.;
\label{eq6}
\end{equation}

\noindent
and a power-law with exponential cutoff (PLEC),

\begin{equation}
\phi(L,z) = L_{n}\left(\frac{L}{L_{break}}\right)^{\alpha}\exp 
\left[\left(\frac{-L}{L_{break}}\right)\right];
\label{eq7}
\end{equation}

\noindent
to model our pulse luminosity function. $\alpha$ and $\beta$ (BPL only) are
the low \& high luminosity indexes; $L_{break}$ is the break luminosity; and
$L_{n}$ is the normalisation of the LF, which is given by the reciprocal of
the LF integral. The normalisation factor is sensitive to the limits of
integration and can have an effect on the derived efficiency parameter,
$K_{GRB}$, up to a factor of 2. The limits of integration are therefore
chosen by various authors depending on the constraints that they place on
their data sets, bias controls, or calculation methods\footnote{The brightest
subsection of low-$z$ GRBs are often utilised as the subsample avoids
Malmquist bias, and is less succeptible to other intrinsic biases such as
redshift detectability, and uncertainty in the CSFRD at
high-$z$ \citep{2011MNRAS.416.2174C}; utilising the least/most luminous
pulses \citep{2006MNRAS.370..185F, 2012ApJ...749...68S}, or integrating over
infinity, especially for PLEC LF models \citep{2010MNRAS.407.1972C}, is also
common.}; the variation in normalisation is small however when compared to the
intrinsic uncertainties in the CSFRD, IMF evolution, metallicity density,
etc.. We adopt the faintest, and brightest pulse luminosities as the limits of
integration, which in this paper spans $1\times10^{46}$ to $1\times10^{54}$ ergs
$s^{-1}$.    

\subsubsection{A Separate Low-Luminosity GRB Population}

Although LGRB studies generally prefer utilising luminosity functions that
assume a single population of LGRBs, a small group of LGRBs appear to exist with
particularly low luminosities (LL, $L<10^{50}$ ergs $s^{-1}$) that are poorly
fitted by these single population models \citep{2007ApJ...662.1111L,
2009MNRAS.392...91V, 2010MNRAS.406..558Q, 2013MNRAS.428..167H}. Typically
these LL LGRBs are assumed to trace the same progenitor models as those of
higher luminosity LGRBs whilst convolved to a separate luminosity function. Such
luminosity functions produce markedly differing normalisation rates for the
two types of LGRBs; the local formation rates of LL LGRBs are suggested to be
several orders of magitude greater than those of more luminous LGRBs.

With the incorporation of bright prompt emission pulses, and late time, faint
X-ray flares, 72 of 607 pulses fall into the luminosity regime typically
associated with LL LGRBs. In this paper we evaluate the performance of bimodal LF models (denoted as type II models) compared to single population LGRB models (type I models). Following a similar procedures set out by \citet{2007ApJ...662.1111L}, we produce a bimodal luminosity function by combining two luminosity functions, $\phi_{LL}(L)$ and $\phi_{HL}(L)$ such that:

\begin{equation}
\phi(L)=\phi_{LL}(L)K_{LL-GRB} +\phi_{HL}(L)K_{HL-GRB},
\label{eq8}
\end{equation}

\noindent
where the LGRB formation rate efficiencies, $K_{LL-GRB}$ and $K_{HL-GRB}$ are
included in the LF to allow for different formation efficiencies of the two
GRB types, and are analogous to the $\rho^{LL}_{0}$ and $\rho^{HL}_{0}$
parameters found in \citet{2007ApJ...662.1111L}. Both $\phi_{LL}(L)$ and
$\phi_{HL}(L)$ follow the same shape as Equations \ref{eq6} and \ref{eq7} and each population is fitted separately to ensure that the LL and HL parameters are independent of each other. Normalisation limits for the bimodal LFs, as that of the single population model, are set at $1\times10^{46}$ to $1\times10^{54}$ ergs $s^{-1}$.  

\subsection{LGRB Co-moving Pulse Rate}

We model the comoving burst rate, $\Psi_{*}(z)$, or more
specifically the comoving pulse formation rate (pulses $yr^{-1}$
$Mpc^{-3}$) using two diametrically opposed models: 

\begin{equation}
\Psi_{*}(z) =  K_{pulse}\left\{
\begin{array}{ll}
K_{GRB}\psi_{*}(z)\iota(z)F(z) & \text{Type I} \\
\psi_{GRB}(z)  & \text{Type III}  \\
\end{array}\right. .
\label{eq9}
\end{equation}

Type I models assume a functional form for the cosmic star formation
rate density (CSFRD), $\psi_{*}(z)$ ($M_{\odot}$ $yr^{-1}$
$Mpc^{-3}$), and couple to: an evolving fraction of high-mass stars that
are capable of forming GRBs at at given redshift, $F(z)$; an additional rate
density evolution parameter, $\iota(z)$, capable of boosting GRB formation
rates above CSFRD levels, and conversion factors $K_{pulse}$, and $K_{GRB}$
which describe the average number of pulses per GRB, and the number of GRBs
formed per solar mass of stars respectively. Included amongst the type I models is a non-evolving GRB luminosity function derived when $\iota(z)$ is constant, and the break luminosity index, $\delta=0$. Type III models are a common alternative to type I models where direct fitting of a simple functional form to $\Psi_{*}(z)$, in this paper taken to be a triple broken power-law, allows for ease of comparison between cosmic star formation rate density models without the need for refitting of GRB luminosity functions. 

All the parameters used in modelling the comoving pulse formation
rate are functions of redshift with the exception of $K_{pulse}$: the number of pulses per GRB shows no correlation with redshift; having removed the effect of the BAT rest-frame duration, $T_{90}/(z+1)$, we derive a Spearmann's partial rank correlation coefficient of $\rho_{s}=-0.045$, implying that $K_{pulse}$ is redshift-independent.    

\subsection{Cosmic Star Formation Rate Density}

The comoving burst formation rate is dependant on the properties of the
central engines that power GRBs; for LGRBs the preferred mechanism is that of
a collapsar: massive stars that undergo catastrophic core collapse into
blackholes \citep{1993ApJ...405..273W, 1998ApJ...494L..45P,
1999ApJ...524..262M}, favoured because of the observed association with Type
Ib/c supernovae \citep{1998Natur.395..670G, 2003ApJ...591L..17S} with Wolf-Rayet
stars the favoured progenitor type. With their high mass ($M > 25M_{\odot}$),
and subsequently short main-sequence lifespans, Wolf-Rayet stars closely trace
the local star formation rate; as such, for type I/II models, we take the
Cole \citep{MNR:MNR4591} functional form for the CSFRD: 

\begin{equation}
\psi_{*}(z)= \frac{(a_{1}+a_{2}z)h}{1+(z/a_{3})^{a_{4}}},
\label{eq10}
\end{equation}

\noindent
in units of $\:M_{\odot}\:yr^{-1}\:Mpc^{-3}$; and use the best fit
parameters: $a_{1}=0.0389$, $a_{2}=0.0545$, $a_{3}=2.973$, and $a_{4}=3.655$
derived by \cite{2013ApJ...763....3K}. These values are based on corrections
to the work by \cite{2006ApJ...651..142H} where overestimations in the CSFRD
were found to have arisen due to uncertainties in the correction for
dust-obscuration and the conversion from UV luminosity to intrinsic star
formation rates. These coefficients produce a cosmic star formation rate that
has an almost flat profile to a redshift of $z=2$ and approximately an order
of magnitude greater formation rate at $z=0$ than that produced from using
Hopkins and Beacom's fitted parameters.

\subsection{The Cosmic IMF}

A contributing second-order effect from an evolving population of high-mass
stars is considered by some authors either explicitly in the modelling of
derived GRB luminosity functions \citep{2002ApJ...574..554L} or as an
explanation to the observed evolution in luminosity or rate
parameters \citep{2008ApJ...673L.119K, 2011MNRAS.416.2174C}. The CSFRD is,
by definition, the total star formation rate at a given redshift and, for
completeness, in this paper we explicitly convert the CSFRD to a formation
rate density of stars capable of undergoing catastrophic core collapse and
forming GRBs (i.e. with mass greater than $25M_{\odot}$) by deriving the
fractional mass of stars greater than a "GRB ignition mass", $F(z)$, given by
$F(z)=\int_{25M_{\odot}}^{120M_{\odot}}M\Phi(M,z)dM/\int_{0.01M_{\odot}}^{120M_{\odot}}M\Phi(M,z)dM$. In our derivation of the fraction of high-mass stars we assume an IMF,
$\Phi(M,z)$, which is top-heavy at high redshift as logically in the
metal-poor early universe the Eddington limit, and subsequently the population
of high mass stars, was much greater than more recent epochs. Studies into
extra-galactic star formation history indicates an evolving
IMF \citep{2008MNRAS.385..147D, 2008ApJ...674...29V, 2008MNRAS.385..687W} up
to $z\sim2$ and as such we adopt the redshift-dependent IMF model
of \cite{2008MNRAS.385..147D} where the IMF takes the form of a broken
power-law \citep{2001ASPC..228..187K}:

\begin{equation}
\Phi(M,z) = \left\{
\begin{array}{ll}
\left(\frac{M}{\hat{M}}\right)^{-0.3} & M \leq \hat{M} \\
\left(\frac{M}{\hat{M}}\right)^{-1.3} & M > \hat{M}  \\
\end{array}\right.
\label{eq11}
\end{equation}

\noindent
with the characteristic break mass evolving with redshift: $\hat{M} =
0.5(z+1)^{2}M_{\odot}$, which we naively extrapolate up to $z=10$. The effect
of the evolving IMF on the distribution of stellar masses is subtle; in the
current epoch, approx. $9.6\%$ of all stellar mass formed per year is locked up
within stars of $M >25M_{\odot}$, increasing to approx. $62.5\%$ at $z=10$.

\subsection{{\em Swift} Detection Likelihood}

\begin{figure}
\centering
\includegraphics[width=\linewidth, bb = 0 0 550 400]{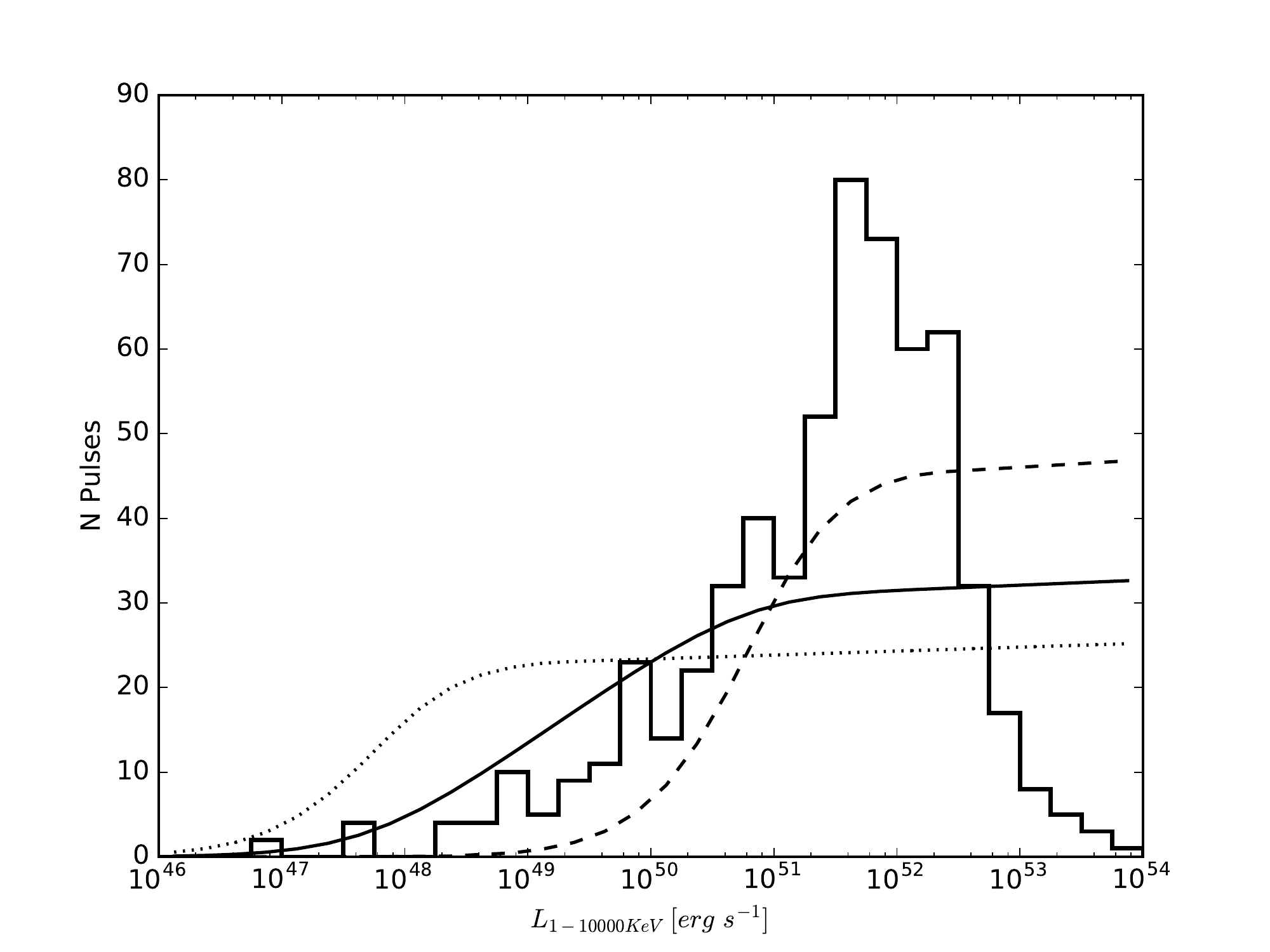}\\
\caption{The observed distribution of pulse
    luminosities for a type I model with rate evolution of $\gamma=0$ where the luminosity function is assumed to be constant. The solid curve is the expected distribution resulting from integrating Equation \ref{eq5} whilst assuming a detection profile, $D(L,z)$, that varies from unity at the BAT threshold down to
    non-detection at the approximate XRT threshold. The dotted curve is the
    expected distribution assuming $D(L,z)$ is unity down to the XRT
    sensitivity limit and the dashed curve arises when setting the BAT
    detection threshold as the lower limit of detection.}
\label{fig4} 
\end{figure}

It is common, in previous studies of the {\em Swift} GRB luminosity function,
where only the defining pulse luminosities (i.e. the brightest) were utilised,
to set the likelihood of detection by the BAT within its field of view to be at
unity. In deriving a GRB pulse luminosity function incorporating data from the
XRT we include pulses up to three orders of magnitude less luminous than the
detection threshold of the BAT. We produce a model of the {\em Swift}
detection profile, $D(L, z)$, assuming total detection likelihood above the BAT
detection threshold which scales to zero at an effective XRT detection
threshold of $3\times10^{-12}$ ergs $s^{-1}$ $cm^{-2}$ as a
power-law of index $\sim-1/3$. This is, of course, a naive model of {\em
Swift's} detection profile: each pulse is treated as an individual event and
assuming unity down to the XRT detection threshold would be inappropriate;
each pulse detected by the XRT was because of BAT detection and the XRT
detection threshold varies considerably from burst to burst depending on the
brightness of the afterglow component, and the time between XRT detection and
BAT trigger; furthermore as there is often significant overlap between pulses,
fainter pulses may be seen when an earlier, significantly brighter pulse is
present. Modelling the combinded detection profile of {\em Swift} is highly
complicated and, as such, the results are somewhat subjective. Our detection
profile convolved to the CSFRD, metallicity density, and constant 
$\phi(L,z)$, produces a distribution of pulses that closely traces the
observed distribution up to approximately $10^{51}$ ergs $s^{-1}$ (solid line,
Figure \ref{fig4}). Setting the detection profile to unity above
either the XRT or BAT detection thresholds produces the dotted and dashed
distributions which tends to overestimate the population of low luminosity
pulses ($>$XRT = unity) or underestimates the population of sub-peak
luminosities ($<$BAT = 0) requiring, respectively, a luminosity function that is
more positively or negatively tilted to compensate.   

\subsection{Redshift Evolution Models}

For a type I GRB LF model, the basic method of taking a CSFRD
convolved to a luminosity function, detection profile, and cosmological volume
element produces a distribution of LGRBs that under-represents the observed
high-redshift, high-luminosity population. The solution is to provide an extra
evolutionary effect in the modelling and allow it to float when fitting the
model parameters. In this paper we look at three of the most common
evolutionary effects: evolution of the break, or cutoff, of the luminosity
function; a metallicity density evolution such that LGRBs trace low
metallicity star forming regions; and a more generic rate density evolution on
top of the CSFRD as solutions to differences between the observered and type I pulse distribution functions.  
 
\subsubsection{Break Luminosity}
Evolution in the break, or cutoff, luminosity is of the form
$L_{break}(z)=L_{0}(z+1)^{\delta}$, where $L_{0}$ is the break in the LF at
$z=0$ and $\delta$ is the index of LF evolution \citep{2002ApJ...574..554L,
2004ApJ...611.1033F, 2004ApJ...609..935Y, 2006ApJ...642..371K,
2007ApJ...656L..49S, 2009MNRAS.396..299S, 2010MNRAS.407.1972C,
2011MNRAS.417.3025V, 2012ApJ...749...68S, 2015arXiv150401414P,
2015arXiv150401812Y}. This has the further effect
that the normalisation parameter, $L_{n}$, becomes $L_{n}(z)$. In this paper
the break luminosity evolution can be applied to both the type I and type III LGRB pulse formation rate models. In principle, luminosity break evolution can be incorporated into type II models such that either one, or both, GRB populations see their own luminosity evolution. Given the large number of free parameters, and the small population of low-luminosity pulses, however, we believe that we do not yet have the statistics to draw meaningful conclusions from such a model.

\subsubsection{Metallicity Density}

Extreme mass-loss through stellar winds, a characteristic of high-mass stars,
will prevent the formation of a GRB; if, however, the progenitor has low
metallicity ($Z < 0.1Z_{\odot}$) then the mass-loss rate is severely dampened
and a GRB is able to form \citep{1999ApJ...526..152F,
2006RPPh...69.2259M}. LGRB progenitors should therefore preferentially form in
low-metallicity galaxies at any given redshift. A model of fractional mass
densities belonging to metallicities below metallicity $Z$ at redshift $z$,
$\Sigma(z)$ has been derived by \cite{2006ApJ...638L..63L} from the Schechter
distribution function of galaxy masses and the mass-metallicity relationship
determined from SDSS surveys. The functional form of $\Sigma(z)$ is given by:

\begin{equation}
\Sigma(z)=\frac{\hat{\gamma}[0.84,(Z/Z_{\odot})^{2}10^{0.3z}]}{\Gamma(0.84)}
\label{eq12}
\end{equation}

\noindent
where $\hat{\gamma}$ and $\Gamma$ are the lower incomplete and complete gamma
functions respectively. The metallicity density will always boost
high-redshift GRB formation rates, with the metallicity threshold determining
how rapidly this rate increases; a higher metallicity threshold will produce
a smaller increase in GRB formation with redshift, tending towards no
evolution when $Z/Z_{\odot}\to\infty$ \citep{2007ApJ...656L..49S,
2010MNRAS.406..558Q, 2011MNRAS.417.3025V, 2012ApJ...749...68S}. 

\subsubsection{Rate Density}

Metallicity density evolution acts as a physical explaination to observed
evolution in GRB formation rates, however the formulation of the model relies
on no scatter in the mass-metallicity relationship, and no redshift evolution
in the faint end of the Schechter galaxy mass function and the rate of which the
average galactic metallicity evolves. One may instead use a simple
$(z+1)^{\gamma}$ factor to produce the same effect as metallicity density
evolution with the advantage that rate density also allows for a dampening of
GRB formation rates at high-$z$, something that is impossible for the
formulation of metallicity density to achieve \citep{2006ApJ...642..371K,
2008ApJ...673L.119K, 2009MNRAS.396..299S, 2010MNRAS.406..558Q,
2011MNRAS.416.2174C, 2011MNRAS.417.3025V, 2012ApJ...744...95R,
2012ApJ...749...68S, 2015arXiv150401414P}. This factor is however purely
empirical, which frustrates interpretations of the 
results. Both the metallicity density and rate density evolution are
incorporated into the type I GRB formation rate model, $\Psi_{*}(z)$,
through the $\iota(z)$ term in Equation \ref{eq9}, either singularly or in
combination with each other \citep{2010MNRAS.406..558Q}.   

\subsubsection{Combined Break Luminosity \& Rate Density}

Evolution either in rate density, or break luminosity has been utilised as a
solution to discrepencies between theoretical, and observed LGRB luminosity
functions. Little study has however been made on the performance of more complex
evolutionary models involving evolution in both rate and break luminosity. In
this paper we evaluate the performance of a type I combined rate/break evolutionary
model and compare this model's performance with the more common univariate type I
evolutionary models.

\section{The GRB Pulse Luminosity Function}
\label{s4}
\subsection{The MCMC Simulation}

We bin the observed distribution, $N(L,z)$, by splitting the \npul GRB pulses
into equipopulous redshift bins: $0.125<z\le1.505$, $1.51<z\le2.6$, and
$2.612<z\le 9.4$; we furthermore bin over luminosity to improve statistics at
the high and low luminosity tails of the GRB pulse distribution such that the
$i$th bin is the associated set $J_i\equiv\{j\vert L_{i}^{min}<L<L_{i}^{max},
z_{i}^{min}<z<z_{i}^{max}\}$, with $n(J_i)$ as the total number of pulses in
$J_i$, set at a minimum of 11 pulses: a tradeoff between maximising the total
number of bins, and reducing the fractional Poissonian error component of each
bin. The lower and upper redshift and luminosity limits of each bin are
subsequently trimmed to remove excess "padding" of empty data space with the
resulting bins shown in Figure \ref{fig3}. For a non-trivial model with
parameters, $\hat{\theta}$, a Gaussian minus log-likelihood function can be
constructed using methods outlined by \cite{2005physics..11182D}, giving:

\begin{multline}
-log(\mathcal{L}[\hat{\theta}\vert n(J_i)])
 = \sum_{i=1}^{N}\frac{log(2\pi\sigma_{i}^{2})}{2} \\
 +\sum_{i=1}^{N}\frac{[n(J_i)
-
\int_{L_{i}^{min}}^{L_{i}^{max}}\int_{z_{i}^{min}}^{z_{i}^{max}}\eta(L,z;\hat{\theta})dLdz]^2}
{2\sigma_{i}^{2}}  
\label{eq13}
\end{multline}

\noindent
where $\eta(L,z;\hat{\theta})$ is equivalent to the R.H.S of
Equation \ref{eq5} and the associated squared error of the $i$th
bin is given by $\sigma_i^2 = \sigma_{n(J_i)}^2 + \sigma_{L_i}^2
+ \sigma_{z_i}^2$. The error in $n(J_i)$, $\sigma_{n(J_i)}$ is naively taken
as the standard deviation of a Poissonian distribution with mean, $n(J_i)$,
giving $\sigma_{n(J_i)}=\sqrt{n(J_i)}$. The errors, $\sigma_{L_i}$, and
$\sigma_{z_i}$ are defined as uncertainties in the limits of integration for
each bin. As the bin edges are defined only by the minimal/maximal pulse
luminosities and redshifts contained theirin, assuming a 10\% uncertainty in
the limits of integration gives:  

\begin{multline}
\sigma_{L_i} = \int_{0.9L_{i}^{min}}^{1.1L_{i}^{max}}\int_{z_{i}^{min}}^{z_{i}^{max}}\eta(L,z;\hat{\theta})dLdz \\ 
- \int_{L_{i}^{min}}^{L_{i}^{max}}\int_{z_{i}^{min}}^{z_{i}^{max}}\eta(L,z;\hat{\theta})dLdz; 
\label{eq14}
\end{multline} 

\begin{multline}
\sigma_{z_i} = \int_{L_{i}^{min}}^{L_{i}^{max}}\int_{0.9z_{i}^{min}}^{1.1z_{i}^{max}}\eta(L,z;\hat{\theta})dLdz \\ 
-\int_{L_{i}^{min}}^{L_{i}^{max}}\int_{z_{i}^{min}}^{z_{i}^{max}}\eta(L,z;\hat{\theta})dLdz. 
\label{eq15}
\end{multline} 

\begin{figure}
\begin{center}
\includegraphics[width=\linewidth, bb = 0 0 500 800]{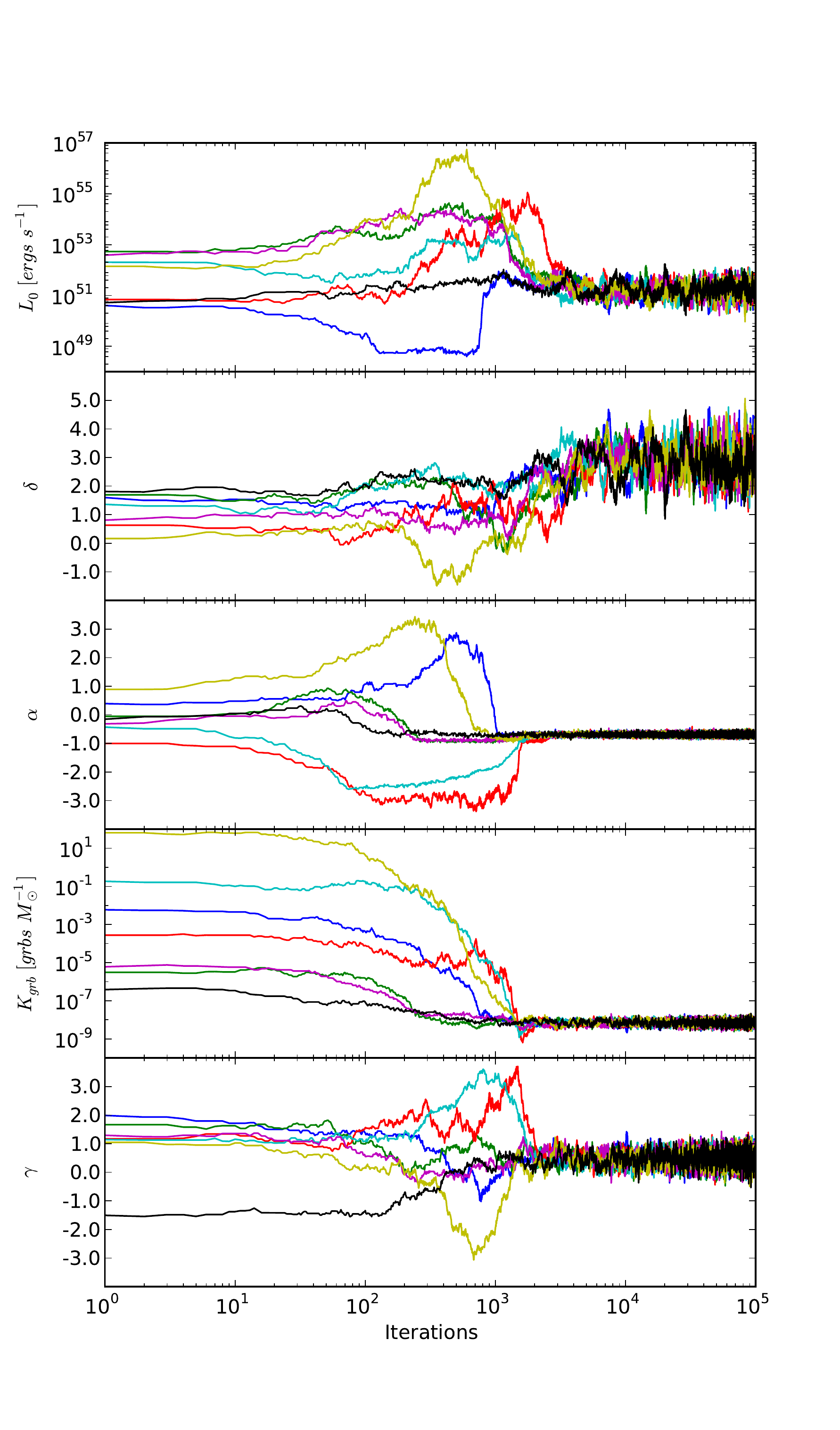}
\end{center}
\caption{7 MCMC chains for the type I GRB pulse LF PLEC model convolved with the
  CSFRD, and the fractional population of high mass stars. The parameters for
  the break luminosity, $L_{0}$; luminosity break evolution, $\delta$;
  low-luminosity index, $\alpha$; GRB formation rate, $K_{grb}$; and GRB
  formation rate evolution, $\gamma$, were allowed to evolve from random start
  points and converge to a unique solution typically around $\sim2\times10^3$
  iterations. Gelman \& Rubin PSRFs statistics for each of the parameters were
  derived with $\sigma_{PSRF}=1.0051,1.0073,1.0058,1.0022,1.0052$ for $L_{0}$,
  $\delta$, $\alpha$, $K_{grb}$, and $\gamma$ respectively.} 
\label{fig5}
\end{figure}

\noindent
A Metropolis-Hastings Markov Chain Monte Carlo (MCMC) method is preferred for
the maximisation of the minus log-likelihood due to the high dimensionality of
the fitting, as well as being able to return the confidence regions of all
fitted parameters. Assuming uniform priors for the indexes: $\alpha$, $\beta$,
$\gamma$, and $\delta$; and logarithmic priors for $K_{GRB}$ and $L_{0}$, we
run MCMCs with chain lengths of $1\times10^{6}$ with typical "burn in"
taking around $2\times10^{3}$ iterations. To ensure that the MCMC program is
finding the global, rather than local, maximum we evaluate the MCMC
convergence success by running multiple MCMCs from random starting points and
deriving the Gelman \& Rubin \citep{Gelman92} potential scale reduction
factors (PSRFs); an example for the fully evolving PLEC model, allowing GRB
rate and break luminosity evolution, is shown in Figure \ref{fig5}. 

For the type I models our results are discussed in Section \ref{s5} and tabulated in Table 1; the results
derived using the type II bimodal low-luminosity and high-luminosity functions
are discussed in Section \ref{s6} and displayed in Table 2; and the results for a type III LF independent of
formation rate models are discussed in Section \ref{s7} and shown in Table 3. The $\chi^{2}_{r}$ quoted are
derived from the 54 bins shown in Figure \ref{fig3}, the associated error of
the $i$th bin, $\sigma_{i}$, and the number of fit parameters of the
model. The Akaike weights, $w_{i}(AIC_{c})$, derived using the Akaike
Information Criterion (AIC) \citep{1974ITAC...19..716A}, are shown and are a
measure of the relative likelihood of each model. Derived from
$w_{i}(AIC_{c})=exp(-\Delta(AIC_{c})_{i}/2)/\Sigma[exp(-\Delta(AIC_{c})_{i}/2)]$
where $\Delta(AIC_{c})_{i}=AIC_{c}-min(AIC_{c})$, they can be considered as
the probability that model $M_{i}$ is the best amongst all the chosen models and
penalises models with larger numbers of free parameters. The $AIC_{c}$ is used
rather than the $AIC$ as it contains extra-terms that adjust for the bias that
a finite sample size can contain.

\section{The Type I GRB Models}
\label{s5}
\subsection{No Evolution Model (Type I-1)}

We find that the scenario in which there is no inclusion of evolutionary models:
luminosity break, rate density, or metallicity, produces a fit of
$\chi^{2}_{r}=[1.81, 1.83]$ for the BPL and PLEC models respectively. This
model produces a distribution of GRB pulses that underestimates the extrema of
the observed pulse luminosity distribution. The derived normalised Akaike
information criterion weights, $w_{i}(AIC_{c})$ for the BPL and PLEC LF models
are $\sim\times10^{-7}$, making these models highly unlikely, compared to the
fully evolving LF and GRB rate type I-5 models, to minimise the Kullback-Leibler
discrepancy and as such we can reject this model. This finding is in agreement
with single pulse studies utilising the brightest prompt emission pulses 
\citep{2006MNRAS.372.1034D, 2007ApJ...656L..49S, 2009MNRAS.396..299S,
  2010MNRAS.406..558Q, 2010MNRAS.406.1944W, 2011MNRAS.417.3025V,
  2012ApJ...749...68S}. 

\subsection{Rate Density Model (Type I-2)}

The addition of a simple $(z+1)^{\gamma}$ rate evolution produced a best fit
to the observed pulse distribution of $\gamma=0.51\pm0.30$ for the BPL and
PLEC models (top left panel, Figure 6). This shifts the
peak of the CSFRD to higher redshifts, boosting the GRB pulse formation rate
at high-$z$ whilst reducing low-$z$ formation rates, producing broadly the
same deficiencies as the non-evolving type I-1 model with regards to reproducing the
observed population of LGRBs at the extrema. A marginal improvement in the
fits of $\chi^{2}_{r}=[1.78, 1.78]$ is seen and the addition of the extra
evolutionary parameter makes this model approximately twice as likely as the
non-evolving type I-1 model to produce our observed GRB pulse distribution according to
Akaike weighting. This is however still approximately $10^{6}$ times less
likely than the fully evolving type I-5 model, making this model highly unlikely and
as such we reject it as a solution to the observed evolution in the GRB pulse
distribution.

Our derived $\gamma$ values are consistent with those derived in single pulse
GRB LF studies, albeit towards the lower end of the distribution
($0.5<\gamma<1.93$, \cite{2008ApJ...673L.119K, 2010MNRAS.406..558Q,
  2011MNRAS.416.2174C, 2011MNRAS.417.3025V, 2012ApJ...744...95R,
  2012ApJ...749...68S, 2014arXiv1412.3969G}). This diversity, in part,
reflects the diversity of GRB formation models used, most notably the CSFRD,
and the selection methods of suitable GRBs preferred by the
authors. Furthermore, excluding the evolving formation rate efficiency
of high-mass stars, $F(z)$, which itself produces a weak rate evolution, would
result in a greater derived $\gamma$ value as such effects are ignored in
other papers. Direct comparisons between studies are difficult given the
variation in methods, and data utilised, however the common result is that
inclusion of a rate density parameter improves the performance of the fit but
is less effective than other evolutionary models (see
\citet{2012ApJ...749...68S} for example).

\begin{figure*}
\begin{center}
\includegraphics[width=14cm, bb = 50 00 600 600]{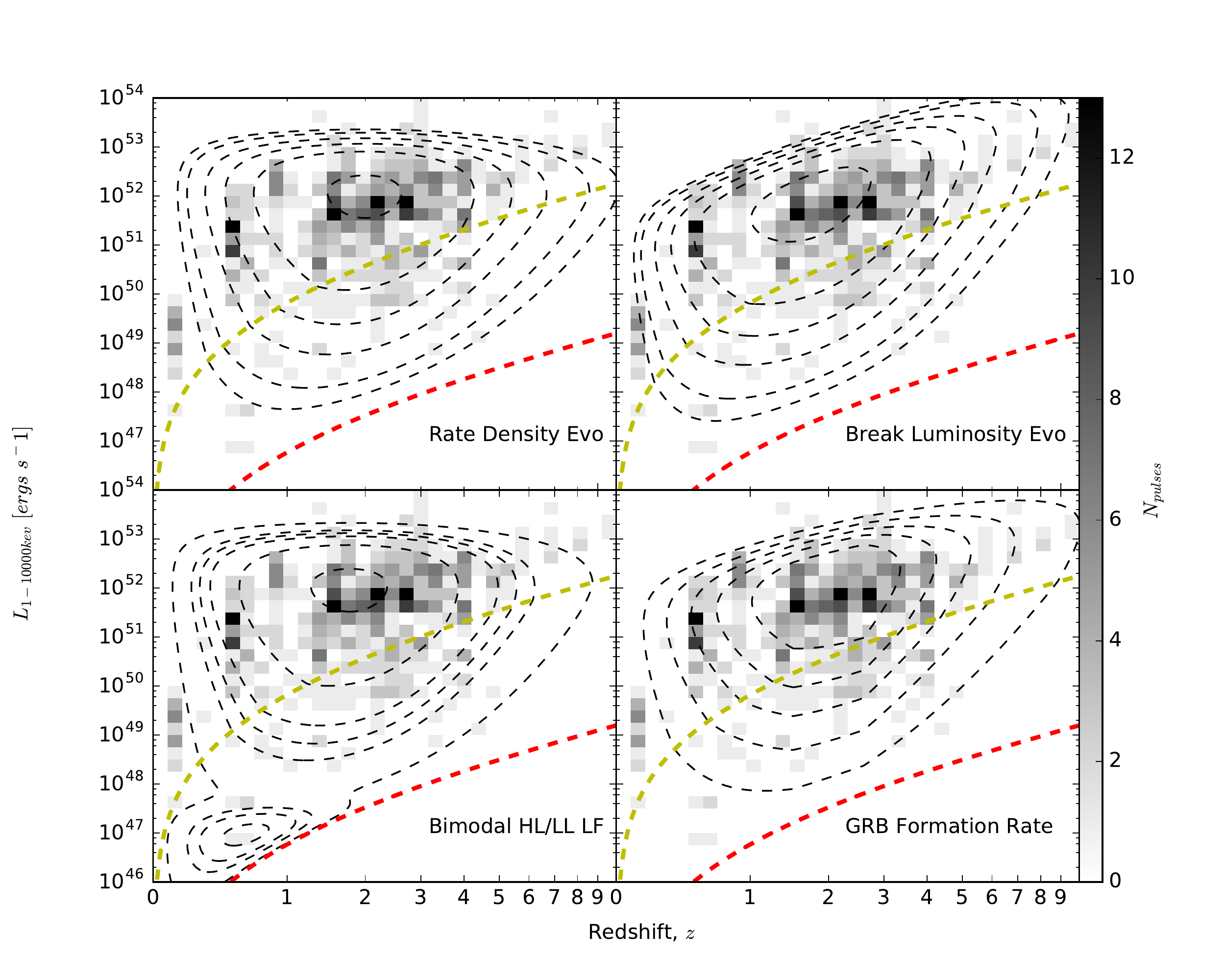}
\caption{The derived GRB pulse luminosity distribution for a PLEC LF (dashed
  contours) overlaying the observed pulse distribution, binned
  to an arbritrary 1/4 dex in luminosity and 1/32 dex in redshift. The four
  models shown corresponds to: a type I-2 rate density evolution model,
  incorporating an additional $(z+1)^{\gamma}$ evolutionary factor (top
  left); a type I-3 break luminosity model evolving as $L_{b}=L_{0}(z+1)^{\delta}$ (top
  right), a type II bimodal GRB luminosity function consisting of separate high and
  low luminosity components (bottom left); and a type III-2 GRB formation rate
  model with break luminosity evolution (bottom right). The upper (yellow) and
  lower (red) dashed lines denote the detection thresholds of the BAT and XRT
  respectively.}
\end{center}
\label{fig6}
\end{figure*}

\begin{table*}
\begin{center}
\begin{tabular}{|l|c|c|c|c|c|c|c|c|c|}
\hline
BPL & $Z/Z_{\odot}$ & $K_{GRB}$ & $\gamma$ & $L_{0}$ & $\delta$ & $\alpha$ & $\beta$ & $\chi^{2}_{r}$ & $w_{i}(AIC_{c})$\\
& & [$10^{-8}\:GRBs\:M_{\odot}^{-1}$] & & [$10^{52}\:ergs\:s^{-1}$] & & & & & \\
\hline
I-1) & - & $1.18^{+0.10}_{-0.09}$ & - & $1.69^{+1.29}_{-0.73}$ & - & $-0.79^{+0.04}_{-0.04}$ & $-1.91^{+2.04}_{-2.04}$ & $1.81$ & $1.4\times10^{-7}$ \\
\hline
I-2) & - & $0.67^{+0.28}_{-0.20}$ & $0.51^{+0.30}_{-0.30}$ & $1.70^{+1.22}_{-0.71}$ & - & $-0.78^{+0.04}_{-0.04}$ & $-1.88^{+1.12}_{-1.12}$ & $1.78$ & $2.6\times10^{-7}$ \\
\hline
I-3) & - & $1.20^{+0.10}_{-0.09}$ & - & $0.02^{+0.05}_{-0.02}$ & $3.35^{+0.74}_{-0.74}$ & $-0.70^{+0.06}_{-0.06}$ & $-1.69^{+0.56}_{-0.56}$ & $1.05$ & $0.0961$ \\
\hline
I-4) & $0.01$ & $653.0^{+66.7}_{-65.6}$ & - & $1.21^{+0.77}_{-0.47}$ & - & $-0.75^{+0.06}_{-0.06}$ & $-1.69^{+0.23}_{-0.23}$ & $1.67$ & $5.1\times10^{-6}$\\
    & $0.1$  & $14.3^{+1.2}_{-1.1}$ & - & $1.05^{+0.74}_{-0.43}$ & - & $-0.75^{+0.06}_{-0.06}$ & $-1.59^{+0.22}_{-0.22}$ & $1.69$ & $4.6\times10^{-6}$ \\
    & $0.2$  & $4.91^{+0.43}_{-0.39}$ & - & $1.05^{+0.77}_{-0.45}$ & - & $-0.75^{+0.06}_{-0.06}$ & $-1.55^{+0.27}_{-0.27}$ & $1.71$ & $4.5\times10^{-6}$\\
    & $0.3$  & $2.87^{+0.25}_{-0.23}$ & - & $1.26^{+0.92}_{-0.53}$ & - & $-0.76^{+0.05}_{-0.05}$ & $-1.60^{+0.43}_{-0.43}$ & $1.71$ & $4.7\times10^{-6}$\\
    & $0.4$  & $2.07^{+0.18}_{-0.17}$ & - & $1.46^{+1.00}_{-0.60}$ & - & $-0.77^{+0.04}_{-0.04}$ & $-1.70^{+0.48}_{-0.48}$ & $1.71$ & $5.6\times10^{-6}$\\
    & $0.5$  & $1.67^{+0.15}_{-0.14}$ & - & $1.67^{+1.04}_{-0.64}$ & - & $-0.77^{+0.04}_{-0.04}$ & $-1.83^{+0.55}_{-0.55}$ & $1.70$ & $6.3\times10^{-6}$\\
    & $0.6$  & $1.47^{+0.13}_{-0.12}$ & - & $1.70^{+1.86}_{-0.89}$ & - & $-0.77^{+0.04}_{-0.04}$ & $-1.89^{+4.33}_{-4.33}$ & $1.71$ & $5.4\times10^{-6}$\\
\hline
I-5) & - & $0.69^{+0.24}_{-0.18}$ & $0.49^{+0.25}_{-0.25}$ & $0.02^{+0.06}_{-0.02}$ & $3.26^{+0.72}_{-0.72}$ & $-0.70^{+0.06}_{-0.06}$ & $-1.67^{+0.72}_{-0.72}$ & $1.00$ & $0.3029$ \\
\hline
\\
\hline
PLEC & $Z/Z_{\odot}$ & $K_{GRB}$ & $\gamma$ & $L_{0}$ & $\delta$ & $\alpha$ & & $\chi^{2}_{r}$ & $w_{i}(AIC_{c})$\\
& & [$10^{-8}\:GRBs\:M_{\odot}^{-1}$] & & [$10^{52}\:ergs\:s^{-1}$] & & & & & \\
\hline
I-1) & - & $1.13^{+0.10}_{-0.10}$ & - & $4.31^{+1.18}_{-0.92}$ & - & $-0.75^{+0.05}_{-0.05}$ & - & $1.83$ & $2.0\times10^{-7}$ \\
\hline
I-2) & - & $0.65^{+0.25}_{-0.18}$ & $0.51^{+0.28}_{-0.28}$ & $4.55^{+1.32}_{-1.02}$ & - & $-0.75^{+0.05}_{-0.05}$ & - & $1.78$ & $4.8\times10^{-7}$ \\
\hline
I-3) & - & $1.19^{+0.10}_{-0.10}$ & - & $0.15^{+0.12}_{-0.07}$ & $2.92^{+0.54}_{-0.54}$ & $-0.70^{+0.05}_{-0.05}$ & - & $1.08$ & $0.0999$ \\
\hline
I-4) & $0.01$ & $642.4^{+58.7}_{-53.8}$ & - & $5.35^{+1.50}_{-1.17}$ & - & $-0.76^{+0.05}_{-0.05}$ & - & $1.79$ & $6.2\times10^{-7}$\\
    & $0.1$  & $13.9^{+1.3}_{-1.2}$ & - & $5.34^{+1.65}_{-1.26}$ & - & $-0.76^{+0.05}_{-0.05}$ & - & $1.80$ & $5.2\times10^{-7}$\\
    & $0.2$  & $4.77^{+0.44}_{-0.40}$ & - & $5.25^{+1.72}_{-1.29}$ & - & $-0.76^{+0.05}_{-0.05}$ & - & $1.79$ & $9.0\times10^{-7}$\\
    & $0.3$  & $2.76^{+0.26}_{-0.23}$ & - & $5.02^{+1.58}_{-1.20}$ & - & $-0.75^{+0.05}_{-0.05}$ & - & $1.75$ & $2.1\times10^{-6}$\\
    & $0.4$  & $1.98^{+0.18}_{-0.17}$ & - & $4.74^{+1.42}_{-1.09}$ & - & $-0.75^{+0.05}_{-0.05}$ & - & $1.73$ & $3.8\times10^{-6}$\\
    & $0.5$  & $1.62^{+0.15}_{-0.14}$ & - & $4.56^{+1.30}_{-1.01}$ & - & $-0.74^{+0.05}_{-0.05}$ & - & $1.72$ & $4.1\times10^{-6}$\\
    & $0.6$  & $1.42^{+0.13}_{-0.12}$ & - & $4.38^{+1.22}_{-0.95}$ & - & $-0.74^{+0.05}_{-0.05}$ & - & $1.72$ & $3.7\times10^{-6}$\\
\hline
I-5) & - & $0.68^{+0.23}_{-0.07}$ & $0.49^{+0.24}_{-0.24}$ & $0.17^{+0.14}_{-0.08}$ & $2.83^{+0.55}_{-0.55}$ & $-0.70^{+0.05}_{-0.05}$ & - & $1.04$ & $0.5010$ \\
\hline
\end{tabular}
\label{tab1}
\caption{Fitted results for Type I models, where the GRB pulse formation rate function, $\Psi_{*}(z)$, incorporates a Cole CSFRD parameterised by \citet{2013ApJ...763....3K}. The best fit parameters for the BPL and PLEC LF models are shown for the scenarios of I-1) where there is no extra evolutionary parameter ($\iota(z)=1$); I-2) there is evolution in the GRB formation rate ($\iota(z)\propto(z+1)^{\gamma}$); I-3) there is evolution in the break luminosity ($L_{break}\propto(z+1)^{\delta}$); I-4) there is present a metallicity density evolution ($\iota(z)=\Sigma(z)$ and $\Sigma(z)=\hat{\Gamma}[0.84, (Z/Z_{\odot})^{2}10^{0.3z}]/\Gamma(0.84)$) and; I-5) there is simultaneously evolution of the rate density and break luminosity ($\iota(z)\propto(z+1)^{\gamma}$ and $L_{break}\propto(z+1)^{\delta}$). The $\chi_{r}^{2}$ values, and Akaike weights, $w_{i}(AIC_{c})$, are derived from the bins shown in Figure \ref{fig4}, and are quoted as a means of comparison rather than the means of fitting. A perfect fit is almost impossible due to the clumpy nature of the pulse data coupled with the affect that uncertainties in the larger bins have on the likelihood.} 
\end{center}
\end{table*}

\subsection{Metallicity Density Model (Type I-4)}

Our attempts at fitting metallicity density evolution proved to be
unsuccessful, with our MCMC code unable to converge on a unique solution,
suggesting strong degeneracy between $Z/Z_{\odot}$ and other fitted
parameters. We therefore chose to set six metallicity thresholds and fit our
data, covering $0.01<Z/Z_{\odot}<0.6$. We find that degeneracy exists between
the metallicity threshold, and all other fitted parameters, with this degeneracy
arising from the unique shape of $\Sigma(z)$. The functional form of
$\Sigma(z)$ can be crudely considered as a linear rise in $z$ connecting two
plateaus at $\Sigma(z)\approx0$ and $\Sigma(z)\approx1$. The metallicity
density threshold acts to shift $\Sigma(z)$ in $z$, whereby a greater
$Z/Z_{\odot}$ shifts the start of the linear rise to lower-$z$. For
$Z/Z_{\odot}=0.01$, this shift is strong enough that the majority of
$\Sigma(z)$ is at the first plateau, resulting in a significantly higher
$K_{GRB}$ to compensate. As $Z/Z_{\odot}$ increases, more of $\Sigma(z)$
occupies the upper plateau and $K_{GRB}$ tends towards values found for type I models
excluding metallicity density evolution. Further degeneracy between
$Z/Z_{\odot}$ and $L_{0}$, $\alpha$, and $\beta$ arises when convolving
$\Sigma(z)$ to the {\em Swift} detection profile, $D(L,z)$. As the detection
thresholds of the BAT and XRT effectively bisects the $L-z$ plane, changes to
the size of the plateau that $\Sigma(z)$ produces is rotated onto the $L$
dimension by $D(L,z)$, and is counterbalanced by variation of the LF
parameters.      

Despite the range of metallicity density thresholds fitted, our type I-4 models
produces broadly similar quality of fits, with small variations as displayed
by the $\chi^{2}_{r}$ and Akaike weights in Table 1. Across all $Z/Z_{\odot}$
in both the BPL and PLEC we see a general improvement in the quality of fits
as compared to both the non-evolving type I-1 model and type I-2 rate density model. The
combined Akaike weights make the type I-4 metallicity density model $153$ times more
likely than the non-evolving type I-1 model and $70$ times more likely than the type I-2 rate
density model whilst the evolving type I-3 LF break model is $3.7\times10^{3}$ more
likely. These values strongly suggests that either the metallicity density
evolution is not a suitable explaination to the observed distribution of GRB
pulses, or that assumptions made in the derivation of $\Sigma(z)$ are not
entirely appropriate. The derivation for the metallicity density evolution by
\cite{2006ApJ...638L..63L} does not, for example, consider scatter in the
mass-metallicity relationship, redshift evolution of the faint end of the
SGMF, or the rate at which the average galactic metallicity evolves with
redshift. 

Although the degeneracies of the metallicity density prevents suitable
convergence in the metallicity density threshold, our results are broadly
similar to studies utilising a GRB's brightest pulse:
\citet{2012ApJ...749...68S} finds that metallicity density evolution is more
likely than rate density evolution and less likely than evolution in the break
to minimise information loss, although to a much less significant
degree than we find; \cite{2010MNRAS.406..558Q} finds that a GRB
formation rate that is proportional to both CSFRD and metallicity density (with
$Z/Z_{\odot}=0.1$) only barely reproduces the $z$ distribution; whilst
\cite{2011MNRAS.417.3025V} finds such models failed to reproduce observations
to enough significance to pass the author's criteria.

\subsection{Break Luminosity Evolution (Type I-3)}

Evolution in the break, or cutoff, of the LF model is the most common
explaination to the observed evolution in the GRB distribution. We find
that the inclusion of break evolution produces an evolutionary factor $\delta$
of $3.35^{+0.74}_{-0.74}$ and $2.92^{+0.54}_{-0.54}$ for the BPL and PLEC
models with corresponding $\chi^{2}_{r}$ values of $1.05$ and $1.08$. As seen
in the top right panel of Figure 6 the evolution in the break
acts to boost the GRB pulse distribution at the extrema, significantly
improving the fit statistics. Combined Akaike weights,
$w_{i}(AIC_{C})$ of $0.196$, for the evolving LF break model shows that this
model is $3\times10^{3}$ times more likely than type I-4 metallicity density models, and
$2.8\times10^{5}$ times more likely than type I-2 rate density models to minimise information
loss; luminosity evolution in the GRB LF, excluding or including all
secondary GRB pulses, is preferred over all over forms of type I univariate
evolutionary models \citep{2007ApJ...656L..49S, 2012ApJ...749...68S}.

Our derived values for the LF break evolution parameter are consistent with
GRB LF studies that utilise {\em Swift} data ($2.1<\delta<3.5$,
\citet{2004ApJ...609..935Y, 2010MNRAS.407.1972C, 2012ApJ...749...68S,
  2015arXiv150401414P, 2015arXiv150605463P, 2015arXiv150401812Y}), whilst
studies that incorporate BATSE data display weaker luminosity evolution
($1.0<\delta<2.0$. \citet{2002ApJ...574..554L, 2004ApJ...611.1033F,
  2006ApJ...642..371K, 2007ApJ...656L..49S, 2009MNRAS.396..299S}). The BPL and
PLEC LF parameters of $\alpha$, $\beta = [-0.70^{+0.06}_{-0.06},
  -1.69^{+0.56}_{-0.56}]$ and $L_{0} = [2.0\times10^{50}, 1.5\times10^{51}]$
ergs $s^{-1}$ are likewise in concordance with those found in the literature,
with shallower low-luminosity gradients generally derived by studies that
incorporate the fainter bursts/pulses detectable by {\em Swift}. 

\subsection{Evolving LF and Rate Density (Type I-5)}

We derive values of $\gamma=0.49^{+0.25}_{-0.25}$ for both LF model types and
$\delta=[3.26^{+0.72}_{-0.72}, 2.83^{+0.55}_{-0.55}]$ for the BPL and PLEC,
with a corresponding $\chi^{2}_{r}$ of $1.00$, and $1.04$ respectively, with
the majority of this improvement is seen in the very high redshift bins. The
derived evolutionary parameters are similar to those of type I-2, I-3 univariate models,
and suggests weak degeneracy between the rate density and break luminosity
model parameters, with the evolution of the break performing the lion's share
of log-likelihood optimisation. The combined Akaike weights makes the type I-5
bivariate evolving model more than $4$ times as likely as the type I-3 evolving LF
break model despite the additional evolutionary parameters required. Although
this suggests that a bivariate evolution model is preferred over a univariate
evolution model, a model based solely on the evolution of the break luminosity
cannot be ruled out entirely.

\subsection{Redshift, Luminosity, and Flux Cuts}

\begin{figure}
\begin{center}
\includegraphics[width=\linewidth, bb = 0 0 500 800]{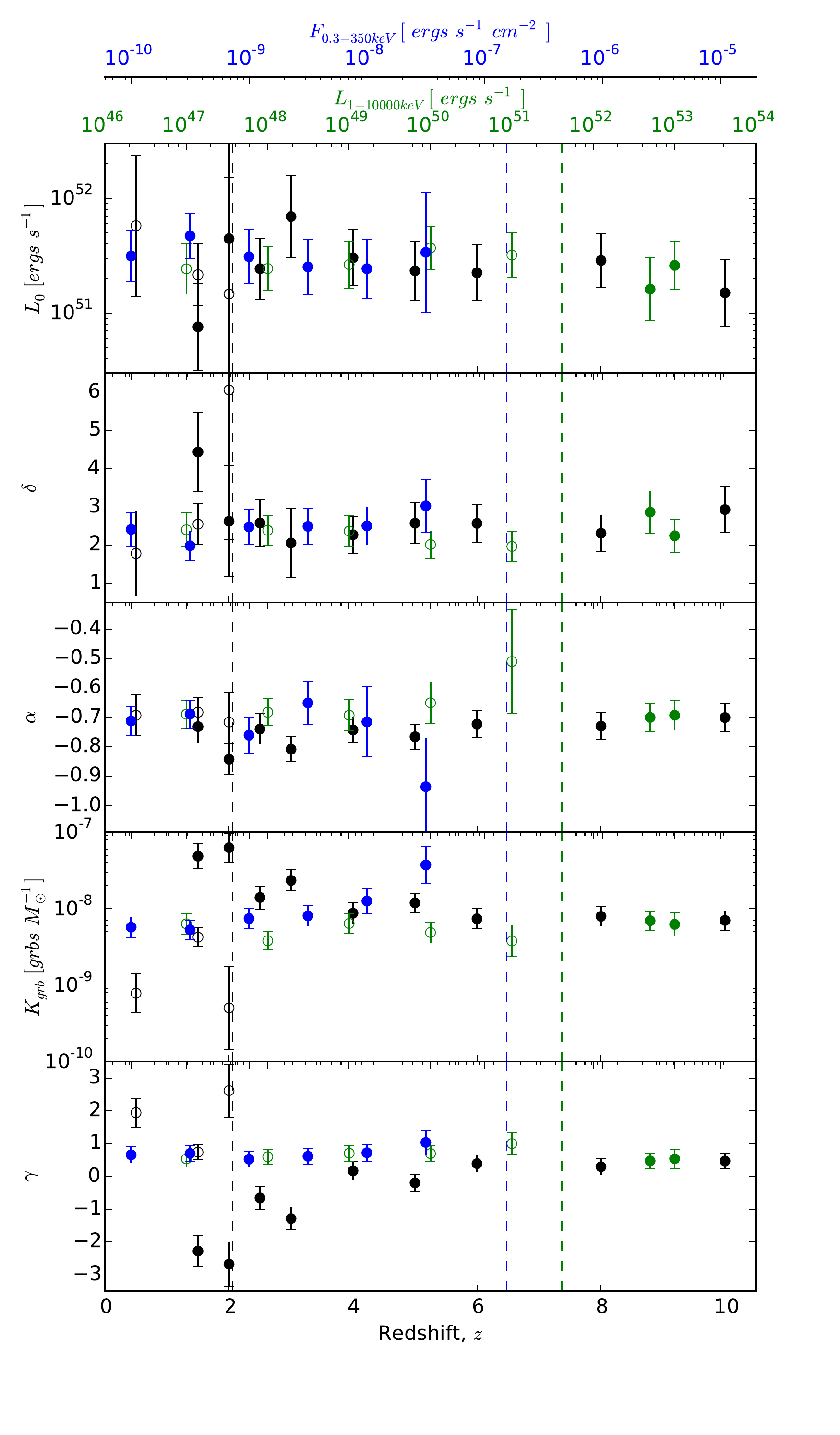}
\end{center}
\caption{Fitted MCMC parameters of a fully evolving type I-5 PLEC LF model
  with successive redshift cuts (black axis), bolometric rest-frame luminosity cuts (green axis), or observer-frame integrated peak flux cuts (blue axis) to the GRB pulse data. Filled and empty circles denote the upper and lower cutoffs respectively for that particular dataset with $1\sigma$ uncertainties. Overlaid are the median redshift, luminosities, and fluxes of the pulse data (dashed lines).} 
\label{fig7}
\end{figure}

In all single population GRB pulse models the residuals of fitted GRB luminosity
functions are greatest at the extrema of the GRB pulse $L-z$
distribution:
\noindent
\begin{itemize}

\item non-evolving models underestimate the population of high-$z$,
high-$L$ pulses, whilst overestimating that of low-$z$, low-$L$ pulses; 

\item rate density models overestimate the population of high-$z$,
high-$L$ pulses, whilst underestimating that of low-$z$, low-$L$ pulses;

\item both models that incorporate break luminosity overestimate the high-$z$,
high-$L$ GRB pulse populations but are consistent with their large associated
uncertainties, contributing little to the log-likelihood function.  

\end{itemize}

\noindent
Discrepencies at the extrema may be due to parent GRBs that are
significantly different from the bulk population, either through a separate
GRB progenitor type (Pop III stars for high-$z$, high-$L$ GRBs) or a via a more
complex GRB luminosity function (LL \& HL GRBs). Cutting away GRB pulses that lie in the extrema of the redshift, or luminosity distributions may produce noticible changes in fitted parameters, suggestive of LGRB
sub-populations. We find, however, that performing successive cuts in the data for the type I-5 PLEC LF model (see Figure \ref{fig7}) of the high/low regions (filled/empty circles) of the $z$, or $L$ distributions (black/green data) produces weak variations in the fitted parameters, which becomes more pronounced as the sample size decreases. Such variations in the fitted parameters are, however, small with good overlap of the $1\sigma$ confidence intervals. Whilst this suggests that the low-$z$/high-$z$, and low-$L$/high-$L$ GRB pulses are part of a single population rather than belonging to unique sub-populations, we cannot
rule out the possibility that the latter is true.

It is common in the data selection phase of GRB luminosity studies to apply a flux cut to the data, with authors arguing that the brightest GRBs in the observer-frame suffer the least from detection bias, and as such are more representative of the true population of GRBs. The study by \citealt{2012ApJ...749...68S}, for example, utilises a flux cut of $P_{15-150 keV}=2.6$ photons $s^{-1}$ $cm^{-2}$ in the observer frame, equating to an integrated flux of $F_{15-150 keV}\approx 1.21 \times 10^{-5}$ ergs $s^{-1}$ $cm^{-2}$ for a PLEC spectrum with $\alpha=-1.57$, $E_{c}=183$ keV. The inclusion of a high flux limit has led to suggestions that the observed evolution seen in such studies arise from a flux threshold selection effect rather than being an intrinsic property of the GRB luminosity function \citep{2013MNRAS.428..167H}. To this end, we vary the flux selection threshold on our GRB pulse data and re-run our MCMCs to refit the data. We find some variation in the GRB pulse luminosity fit parameters (Figure \ref{fig7}, blue data) however such variation, like those of the redshift and luminosity thresholds, are consistent with the intrinsic uncertainties of the model fit parameters. Whilst the direction of the rate evolution parameter, $\gamma$, varies in direction such that it is not concrete that such evolution is real, the evolution in the break luminosity is strong and sees little variation when applying various data cuts.

\section{The Type II GRB Model}
\label{s6}
To reduce the dimensionality of the bimodal LL \& HL GRB LF model, we fix
the indexes of the two populations at the values derived for a Type I-1, non-evolving
single population GRB LF, such that 
$\alpha=\alpha_{HL}=\alpha_{LL}$ (or $\beta=\beta_{HL}=\beta_{LL}$). We find
little variation between the HL LF parameters and the single population Type I-1
LF parameters, unsurprising given the bulk of the GRB pulse population lies
within the regime of HL GRBs. We find a local HL GRB formation rate density,
$\rho^{HL}_{0}$ of $0.22^{+0.02}_{-0.02}$, $0.21^{+0.01}_{-0.02}$ GRBs
$Gpc^{-3}$ $yr^{-1}$ for the PLEC and BPL LFs respectively, compared to the
$0.09<\rho^{HL}_{0}<1.2$ GRBs $Gpc^{-3}$ $yr^{-1}$ range found by
\cite{2007ApJ...662.1111L, 2009MNRAS.392...91V, 2013MNRAS.428..167H} for their
high luminosity GRBs. The inclusion of a secondary LL LF marginally improves
the fitting of the observed $L-z$ GRB pulse distribution (see bottom left
panel, Figure 6), reducing the $\chi^{2}_{r}$ contribution of the low-$L$,
low-$z$ bins at the expense of twice the number of input parameters. The
secondary LL GRB pulse LF shows a break at $L_{0}^{LL}=7.21\times10^{46}$,
$1.10\times10^{47}$ ergs $s^{-1}$ for the BPL and PLEC LF models respectively,
with a local GRB formation rate density of $\rho^{LL}_{0}= 0.09$, $0.21$ GRBs
$Gpc^{-3}$ $yr^{-1}$.   

The ratio of low/high-luminosity GRB formation rate densities found in this
paper are approximately at unity, compared to the ratios of $50-200$ found in
favour of LL GRBs \citep{2007ApJ...662.1111L, 2009MNRAS.392...91V,
  2013MNRAS.428..167H}; varying the limits of normalisation of the LFs has
a small effect on the derived $K_{GRB}$ efficiencies and is not a solution to
the discrepency. Despite a sample of 72 GRB pulses, we are unable to
constrain uncertainties in the fitted parameters. Although reproducing the
observed $L-z$ GRB pulse distribution, the overall combined Akaike weights for
the Type II models ($w_{i}(AIC_{C})=0.1332$), versus the non-evolving Type I-1 LF
models ($w_{i}(AIC_{C})=0.8668$) strongly suggests that a single, non-evolving
population of GRB pulses is a better representation of the $L-z$ distribution
and is not a suitable explanation to the observed evolution of the break
luminosity. We do not rule out that LL GRB pulses are a separate subgroup,
however our data does suggest that it is highly unlikely. 

\begin{table}
\begin{center}
\begin{tabular}{|l|c|c|}
\hline
BPL & Type I-1 & Type II \\
\hline
$K^{HL}_{GRB}$  & $1.18^{+0.10}_{-0.09}\times10^{-8}$ & $1.18^{+0.10}_{-0.09}\times10^{-8}$\\
$L^{HL}_{0}$    & $1.69^{+1.29}_{-0.73}\times10^{52}$ & $1.70^{+0.40}_{-0.32}\times10^{52}$\\
$\alpha^{HL}$   & $-0.79^{+0.04}_{-0.04}$ & $-0.79$ \\
$\beta^{HL}$    & $-1.94^{+2.04}_{-2.04}$ & $-1.94$ \\
$K^{LL}_{GRB}$  & - & $0.50\times10^{-8}$ \\
$L^{LL}_{0}$    & - & $7.21\times10^{46}$ \\
$\alpha^{LL}$   & - & $-0.79$ \\
$\beta^{LL}$    & - & $-1.94$ \\
$\chi^{2}_{r}$  & $1.81$ & $1.91$ \\
$w_{i}(AIC_{C})$& $0.3587$ & $0.0088$ \\
\hline
PLEC & Type I-1 & Type II \\
\hline
$K^{HL}_{GRB}$  & $1.13^{+0.10}_{-0.10}\times10^{-8}$ & $1.12^{+0.10}_{-0.09}\times10^{-8}$ \\
$L^{HL}_{0}$    & $4.31^{+1.18}_{-0.92}\times10^{52}$ &  $4.23^{+0.85}_{-0.71}\times10^{52}$ \\
$\alpha^{HL}$   & $-0.75^{+0.05}_{-0.05}$ & $-0.75$ \\
$K^{LL}_{GRB}$  & - & $1.11\times10^{-8}$ \\
$L^{LL}_{0}$    & - & $1.10\times10^{47}$ \\
$\alpha^{LL}$   & - & $-0.75$ \\
$\chi^{2}_{r}$  & $1.83$ & $1.85$ \\
$w_{i}(AIC_{C})$& $0.5081$ & $0.1244$ \\
\hline
\end{tabular}
\label{tab2}
\caption{The fitted Type I-1, non-evolving model GRB LF for a single population of GRBs (Table 1) and a Type II, bimodal population of high-luminosity ($L>10^{50}$ ergs $s^{-1}$) and low-luminosity ($L<10^{50}$ ergs $s^{-1}$) GRBs. $K_{GRB}$ is given in units of GRBs $M^{-1}_{\odot}$ and $L_{0}$ is in units of ergs $s^{-1}$. Uncertainties in $\alpha$ and $\beta$ for the bimodal population are not given as these parameters were fixed beforehand. The fitted LL GRB LF parameters are quoted without associated errors as they are unconstrained. The $\chi_{r}^{2}$ values, and Akaike weights, $w_{i}(AIC_{c})$, are derived from the bins shown in Figure \ref{fig4}.}
\end{center}
\end{table}

\section{The Type III GRB Models}
\label{s7}
\begin{table}
\begin{center}
\begin{tabular}{|l|c|c|}
\hline
BPL & Type III-1 & Type III-2 \\
\hline
$a_{1}$          & $2.69^{+0.45}_{-0.45}$ & $2.48^{+0.90}_{-0.90}$\\
$z_{1}$          & 1.5 & 1.5 \\
$a_{2}$          & $0.08^{+0.33}_{-0.33}$ & $0.23^{+0.51}_{-0.51}$\\
$z_{2}$          & 2.6 & 2.6 \\
$a_{3}$          & $-1.83^{+1.01}_{-1.01}$ & $-1.97^{+1.43}_{-1.43}$ \\
$\rho_{0}$       & $7.02^{+3.94}_{-2.52}\times10^{-2}$ & $8.29^{+8.13}_{-4.10}\times10^{-2}$ \\
$L_{0}$          & $1.71^{+0.52}_{-0.39}\times10^{52}$ & $0.15^{+0.20}_{-0.09}\times10^{52}$\\
$\delta$         & - & $2.04^{+0.45}_{-0.45}$ \\
$\alpha$    & $-0.72^{+0.12}_{-0.12}$ & $-0.69^{+0.09}_{-0.09}$ \\
$\beta$     & $-1.79^{+0.22}_{-0.22}$ & $-1.88^{+0.25}_{-0.25}$ \\
$\chi^{2}_{r}$   & $2.14$ & $1.43$ \\
$w_{i}(AIC_{C})$ & $1.1\times10^{-7}$ & $0.0415$  \\
\hline
PLEC & Type III-1 & Type III-2 \\
\hline
$a_{1}$          & $2.23^{+1.32}_{-1.32}$ & $2.39^{+1.25}_{-1.25}$\\
$z_{1}$          & 1.5 & 1.5 \\
$a_{2}$          & $0.73^{+0.38}_{-0.38}$ & $0.85^{+0.44}_{-0.44}$\\
$z_{2}$          & 2.6 & 2.6 \\
$a_{3}$          & $-2.20^{+0.95}_{-0.95}$ & $-2.27^{+0.82}_{-0.82}$ \\
$\rho_{0}$       & $7.98^{+15.19}_{-5.23}\times10^{-2}$ & $7.42^{+14.06}_{-4.86}\times10^{-2}$ \\

$L_{0}$          & $4.02^{+1.25}_{-0.96}\times10^{52}$ & $0.36^{+0.48}_{-0.21}\times10^{52}$\\
$\delta$         & - & $2.06^{+0.70}_{-0.70}$ \\
$\alpha$    & $-0.71^{+0.07}_{-0.07}$ & $-0.66^{+0.06}_{-0.06}$ \\
$\chi^{2}_{r}$   & $2.13$ & $1.42$ \\
$w_{i}(AIC_{C})$ & $5.5\times10^{-6}$ & $0.9585$  \\
\hline
\end{tabular}
\label{tab3}
\caption{The fitted parameters for the Type III GRB LF models using a triple broken power-law function that is directly fitted to the GRB formation rate, either excluding (Type III-1) or including (Type III-2) evolution of the LF break. $a_{1}$, $a_{2}$, and $a{3}$ are the gradients of the three power-laws; $z_{1}$, and $z_{2}$ are the redshift breaks, set at the bin edges discussed in Section 4.1; and $\rho_{0}$ is the GRB formation rate density at redshift $z=0$. $L_{0}$ is given in units of ergs $s^{-1}$, and $\rho_{0}$ in units of GRBs $Gpc^{-3}$ $yr^{-1}$.}
\end{center}
\end{table}

\subsection{No Evolution Model (Type III-1)}

We find that our fit utilising the Type III-1 GRB formation rate model, with $\chi^{2}_{r}$
values of $2.14$, and $2.13$, produces a strong rise in GRB pulse formation
rates from the current epoch, plateauing at $z=1.5$, before decaying strongly
away at $z=2.6$. This follows a similar shape as the CSFRD and produces similar
$L_{0}$, $\alpha$, and $\beta$ values as the equivalent Type I/II models
fitted in Section 5, and 6. On initial inspection the Type III-1 models perform less well in fitting, as it requires more than twice the number of model parameters to achieve similar likelihoods, and suffers from the
same inability to reproduce GRB pulse formation rates at the extrema of the
$L-z$ pulse distribution as that of extra rate evolution for Type I-2 models. Whilst this would lead to one assuming that a phemonenological model is better than an empirical one, it is important to note that the CSFRD models have significant uncertainties which are almost universally ignored when propagating errors, creating a false impression of greater quality; it is for this reason that we do not cast any favourable opinion on Type I over Type III models. The combined Akaike weight of $AIC_{c}=5.61\times10^{-6}$ for both non-evolving type III-1 LF models reinforces the conclusion that evolution in the break
luminosity are required to reproduce the observed pulse $L-z$ distribution. This becomes more clear when looking at the probability, and cumulative density functions (Figure 8); the dashed lines, corresponding to a
non-evolving type III-1 model produces a CDF that fails to reproduce the clear
luminosity evolution seen across the three redshift bins, with distinct
underestimation of the luminosities of high-$z$ pulses, and overestimation of
the luminosities of low-$z$ pulses.     

\begin{figure*}
\begin{center}
\includegraphics[width=14cm, bb = 0 0 600 900]{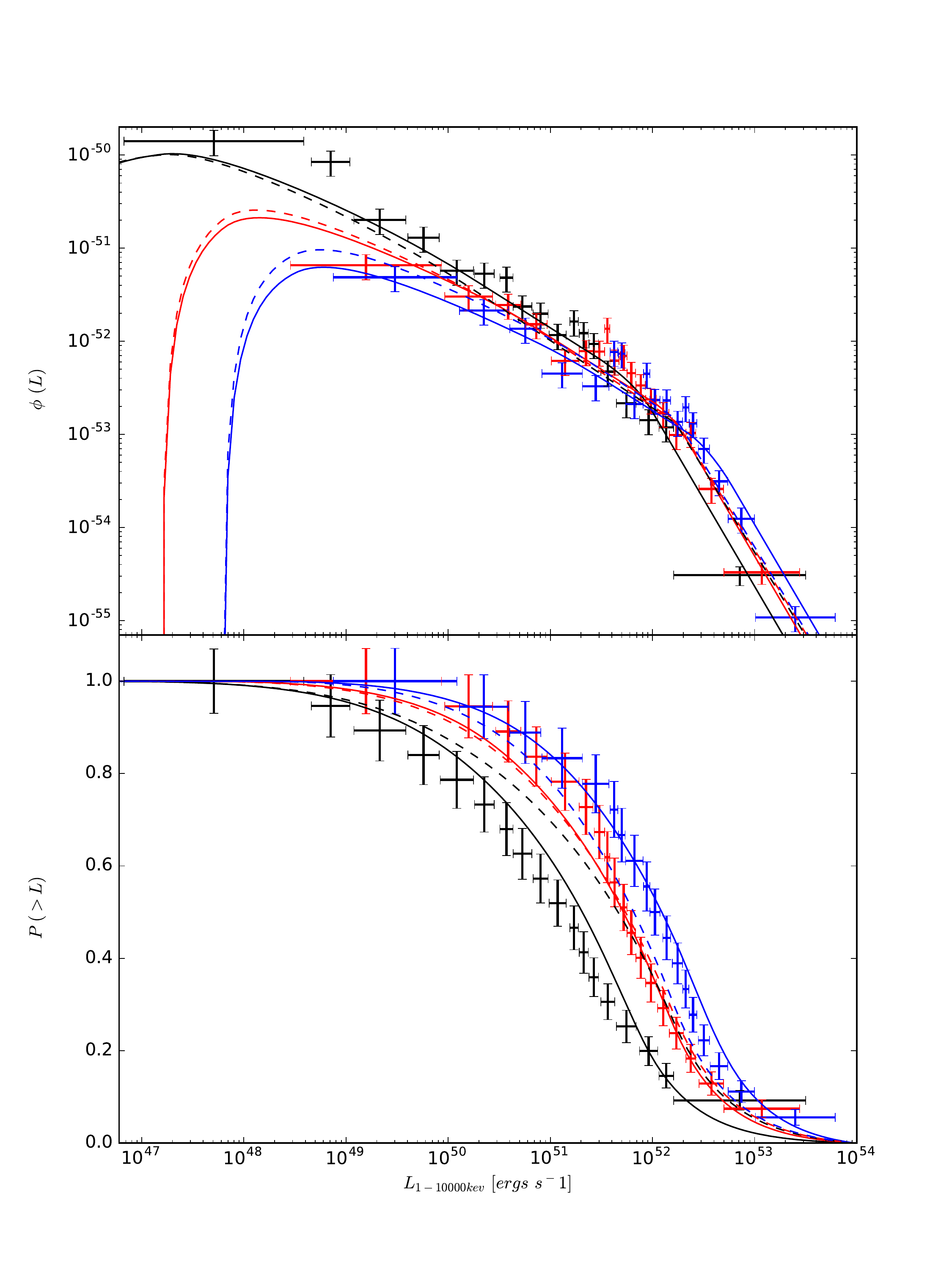}
\caption{The observed, and derived GRB pulse probability densities (top) for
  the three redshift bins, and the equivalent cumulative probability densities
  (bottom) for a type III GRB pulse formation models with a broken power-law
  LPDF. Black data, and lines correspond to the $0.125< z \leq 1.505$
  whilst red, and blue data correspond to the $1.51< z \leq 2.6$ and
  $2.612< z \leq 9.4$ bins respectively. The solid lines denote the type III-2
  model with an additional $(z+1)^{\delta}$ evolution in the break of the
  LPDF, whilst the dashed line is a non-evolving type III-1 model. The turn-off at low luminosities is due to the convolution to the {\em Swift} detection profile.}
\end{center}
\label{fig8}
\end{figure*}

Our pulse luminosity function is consistent with other studies that fit a triple
power-law to the GRB formation rate. Although we utilise all pulses within a GRB
lightcurve and our redshift breaks in the GRB formation rate differ, we find
good agreement with the low-redshift, and high-redshift indexes of
\cite{2010ApJ...711..495B} (BBP), and \cite{2010MNRAS.406.1944W} (WP). We
derive an $\alpha_{1}$, and $\alpha_{3}$ of $2.69^{+0.45}_{-0.45}$,
$-1.83^{+1.01}_{-1.01}$ respectively, compared to: $3.35^{+0.74}_{-0.74}$,
$-2.51^{+1.60}_{-2.25}$ (BBP); and $3.1^{+0.7}_{-0.7}$, $-2.9^{+1.6}_{-2.4}$
(WP) for their models that include GRBs with known redshifts. The
intermediate-redshift indexes, $\alpha_{2}$, derived by those studies
($1.32^{+0.58}_{-0.58}$, BBP; and $1.4^{+0.6}_{-0.6}$, WP) are significantly
stronger than the $0.08^{+0.33}_{-0.33}$ we find, and can be explained by the
difference in position of the first redshift break those authors utilise, who,
like ourselves, do not set as a free parameter in their fitting. 

Our luminosity functions produce a stronger low-luminosity index than these
studies, possibly due to the large number of low-luminosity BAT and XRT pulses
we incorporate, with $\alpha=-0.72^{+0.12}_{-0.12}$ compared to
$-0.27^{+0.19}_{-0.19}$ (BBP), and $0.22^{+0.18}_{-0.31}$ (WP). Our
low-luminosity index is however consistent both with our Type I and Type II models, and
other studies that utilise a CSFRD. The high-luminosity index, $\beta$,
derived by \cite{2010ApJ...711..495B} at $-3.46^{+1.53}_{-1.53}$ is
significantly stronger than our own derived results of
$-1.79^{+0.22}_{-0.22}$; however our results are consistent with
$-1.4^{+0.3}_{-0.6}$ of \cite{2010MNRAS.406.1944W} and is most likely due to
both our studies utilising peak luminosities rather than the time-averaged
luminosities used by \cite{2010ApJ...711..495B}. The break luminosity,
$L_{0}$, derived in this paper at $10^{52.23^{+0.12}_{-0.12}}$ ergs $s^{-1}$
is lower than either studies finds and is consistent with
$10^{52.5^{+0.2}_{-0.2}}$ ergs $s^{-1}$ (WP) but not with
$10^{52.95^{+0.31}_{-0.31}}$ (BBP). 

\subsection{Break Luminosity Evolution (Type III-2)}

Like the Type I-3 model, inclusion of evolution in the break of the luminosity function
significantly improves the quality of fits ($\chi^{2}_{r}=1.43$, $1.42$
for the BPL, and PLEC respectively), with improvement at the
high-$L$, high-$z$ extrema of the pulse $L-z$ distribution (bottom right
panel, Figure 6). With a combined Akaike weight of $w_{i}(AIC_{C})\sim1$ for
the BPL, and PLEC models, the evolving Type III-2 luminosity functions are
$1.7\times10^{5}$ times more likely than the non-evolving Type III-1 models to reproduce the observed pulse distribution. As shown in Figure 8, the CDF for a BPL LPDF model with break luminosity evolution is able to reproduce the observed CDFs for all redshift bins, within $1\sigma$ uncertainties. The evolutionary index of the break luminosity, at
$\delta=2.04^{+0.45}_{-0.45}$, and $2.06^{+0.7}_{-0.7}$ for the BPL, and PLEC
models is weaker than that found using the fully evolving Type I-5 model but
remains consistent with other {\em Swift} studies (see Section 5.4 for
references).

\section{LGRBs and the CSFRD}
\label{s8}
\begin{figure*}
\begin{center}
\includegraphics[width=12cm, bb = 0 0 600 900]{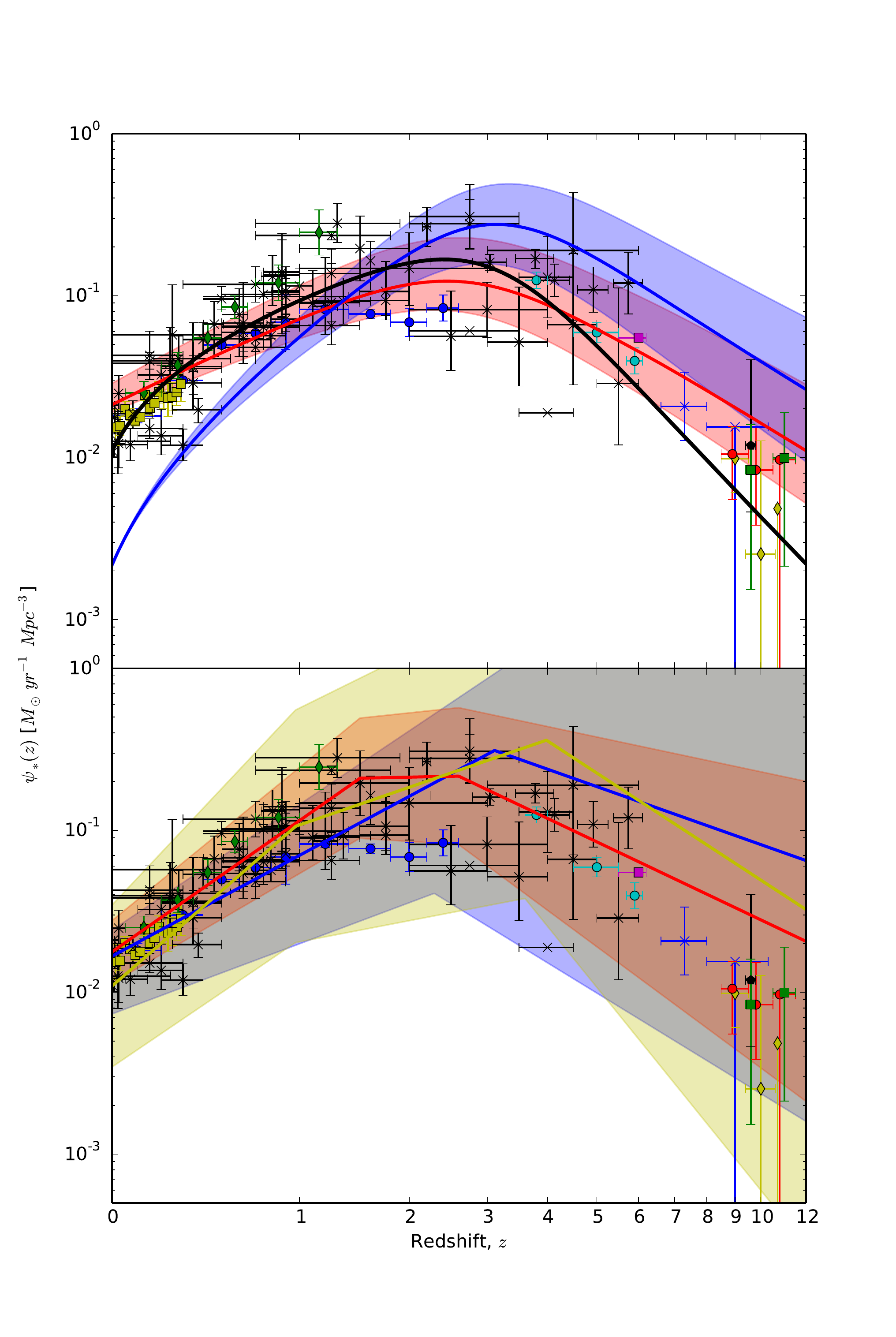}
\end{center}
\caption{Observed cosmic star formation rate densities normalised to the results
  collated by \citet{2006ApJ...651..142H} (solid black line best fit) with
  data from \citet{2003ApJ...595..589B} (magenta squares),
  \citet{2006ApJ...651..142H} (black crosses), \citet{2005MNRAS.358..441B}
  (yellow squares), \citet{2005ApJ...630...82P} (blue circles),
  \citet{2007ApJ...670..928B} (cyan circles), \citet{2008ApJ...686..230B}
  (blue crosses), \citet{2010ApJ...718.1171R} (green diamonds),
  \citet{2012Natur.489..406Z} (black pentagons), \citet{2013ApJ...762...32C}
  (green squares), \citet{2013ApJ...773...75O} (yellow diamonds), and
  \citet{2013ApJ...763L...7E} (red circles). Overplotted for comparison are
  the derived GRB formation rate densities using the rate evolution GRB LF models
  (top panel) of \citet{2012ApJ...749...68S} (equivalent to a Type I-2 model; PLEC LF, blue line), and the
  rate \& break luminosity evolution model derived in this work (Type I-5 model; PLEC LF, red
  line); and the direct GRB formation rate fitted LF models (bottom panel), equivalent to a Type III-1 model, of
  \citet{2010ApJ...711..495B} (yellow line), \citet{2010MNRAS.406.1944W} (blue
  line), and the non-evolving Type III-1 model of this work (BPL LF, red
  line). Shaded regions denote $1\sigma$ uncertainties. All GRB formation
  rates are are normalised to the observed star formation rates by
  maximising a minus log-likelihood function, producing normalisation
  constants, in units of $M_{\odot}$ $GRB^{-1}$, of $6.56\times10^{9}$
  \citep{2012ApJ...749...68S}, and $0.17\times10^{9}$ (this work, Type I-5 PLEC LF model with rate \& break luminosity evolution); and $2.74\times10^{9}$ \citep{2010ApJ...711..495B}, $1.34\times10^{7}$ \citep{2010MNRAS.406.1944W}, and $2.53\times10^{8}$ (this
  work, Type III-1 BPL LF model).} 
\label{fig9}
\end{figure*}

The fitted GRB formation rate densities for Type I-5 and Type III-1 models
derived in this paper are shown in Figure \ref{fig9} overlain with observed
cosmic star formation rate densities, the parameterised CSFRD model of
\citet{2006ApJ...651..142H}, and GRB formation rate models of
\citet{2012ApJ...749...68S, 2010ApJ...711..495B,
  2010MNRAS.406.1944W}. Normalised to the CSFRD, our Type I-5 (including
rate and luminosity evolution) and Type III-1 models trace the observed CSFRD well,
especially at low/intermediate redshifts ($z<5$), with up to a factor 2
deviation between derived high-$z$ ($z>5$) pulse rates and cosmic star formation
rates for the Type III-1 model; suggestive of a redshift break at an earlier epoch
than that which was assumed.

The parameterised model of \citet{2006ApJ...651..142H}, common in the GRB
literature as a model for CSFRD, traces the CSRFD at low redshifts, with
noticible drop-off at high-redshifts. Assuming that the GRB formation rate
follows the CSFRD only, requires the addition of GRB formation rate evolution
to the Type I-1 models to boost high-$z$ GRB formation rates. This addition may,
however, suggest that the parameterised CSFRD models are incorrect at high-$z$
rather than implying the rate of GRB formation was greater at earlier
epochs. Comparing the performance of our rate density evolving Type I-2 model, and the CSFRD as parameterised by \citet{2006ApJ...651..142H} to the
observed CSFRD shows that our Type I-2 model with rate density produces a
log-likelihood of $106.67$ compared to a log-likelihood of $-17.78$
for \citet{2006ApJ...651..142H}. The corresponding Akaike weights for the Type I-2 model is $\sim1.0$\footnote{The log-likelihood for the Type I-2 model is calculated after the unknown conversion factor from
GRB formation rate density to CSFRD is accounted for, and as such the
comparison is with regards to the shape of the CSFRD.}, indicating that the
GRB formation rate evolution seen is not real but is, instead, an artifact of
utilising inappropriate CSFRD models. Utilising GRB formation rates as a probe
to high-$z$ star-formation is therefore highly speculative: conversion from
GRB formation rates to star-formation rates are often cyclical; a
star-formation rate and GRB evolution rate are assumed in order to derive a
GRB formation rate, with which a star-formation rate is
derived \citep{2008ApJ...673L.119K, 2012ApJ...744...95R}; as such, careful
consideration is required when attempting to derive CSFRD models using
high-$z$ burst rates.   

\subsection{The Local GRB Formation rate}

The GRB formation efficiency parameter, in combination with the star formation
rate density at $z=0$, produces the local GRB formation rate density,
$\rho_{0}$. For the Type I models excluding rate density or metallicity
density evolution (I-1, I-3), the formation efficiency, $K_{GRB}$, was
derived to be $K_{GRB}=1.18^{+0.10}_{-0.10}\times10^{-8}$ GRBs $M^{-1}_{\odot}$, in good
agreement to the values of $1.07\pm0.11\times10^{-8}$ and
$1.05\pm0.05\times10^{-8}$ GRBs $M_{\odot}^{-1}$ \citep{2007ApJ...656L..49S,
  2009MNRAS.396..299S}. This equates to a local formation rate density of
$\rho_{0}=0.22^{+0.02}_{-0.02}$ GRBs $Gpc^{-3}$ $yr^{-1}$; for 
models including rate density (I-2, I-4) this drops to
$\rho_{0}=0.12^{+0.05}_{-0.04}$ GRBs $Gpc^{-3}$ $yr^{-1}$. For the Type III
models, the local GRB formation rate is one of the model parameters, and for
a non-evolving BPL LF model this produces a $\rho_{0}$ of
$0.07^{+0.04}_{-0.03}$ GRBs $Gpc^{-3}$ $yr^{-1}$, increasing to $0.083^{+0.08}_{-0.04}$
GRBs $Gpc^{-3}$ $yr^{-1}$ for a Type III-2 evolving BPL LF model. These values are
towards the lower end of the distribution of values found in the literature
for models excluding jet-beaming ($0.03<\rho_{0}<7.3$ GRBs $Gpc^{-3}$
$yr^{-1}$ \citet{2001ApJ...548..522P, 2001ApJ...552...36S,
  2005ApJ...619..412G, 2010MNRAS.406.1944W, 2011MNRAS.416.2174C,
  2012ApJ...749...68S, 2015arXiv150401812Y}).

\section{Conclusions}
\label{s9}
The lightcurves of Gamma-ray bursts exhibit wide variation in temporal
fluctuations with some showing single, bright FRED-like profiles whilst others
have multiple peaks, often with significant overlap. Utilising a physically
motivated model (Willingale et al. 2010) that considers the entire prompt, and
late-time emission as a series of pulses and/or flares with corresponding
emission spectra modelled by a power-law with exponential cutoff, and in most
cases an afterglow component, we fit 118 LGRBs covering the period from
26/01/05 - 03/05/11. This produced 607 GRB pulses spanning $10^{46}<L<10^{54}$ 
ergs $s^{-1}$ in bolometric luminosity, with known redshifts up to
$z\sim10$. 

Traditionally, the brightest pulse of a GRB with known redshift is used as the
defining luminosity of the burst. Such pulses however do not exhibit any other
unique quality: they are often not the first pulse to trigger the BAT, nor do
they solely occur within the prompt emission; they do not possess the hardest
spectrum within a lightcurve, nor do they see the greatest hard-to-soft
evolution of said spectrum; even their brightness is, in some cases, hardly
unique as some bursts contain multiple pulses of comparable brightness. A
great deal of information is therefore lost when utilising solely the
brightest pulses, compounding the difficulties in population analysis for a
relatively rare phenomena which, until recently, was sparcely populated. We
therefore choose to compute the GRB pulse luminosity function, of which the
traditional GRB luminosity function can be considered as either a high-luminosity, or high-flux
sub-population. We convolve a GRB pulse luminosity probability density
function to a GRB formation rate model using three of the most popular GRB LF
theories in the literature: Type I models that traces the cosmic star
formation rate, convolved with various evolutionary effects such as break
luminosity (I-3), rate density (I-2) and metallicity density evolution (I-4); Type II models that are bimodal in nature, allowing for distinct populations of low-luminosity and high-luminosity GRB pulses; and Type III models that are fitted directly to the GRB formation rate. We consider both
PLEC, and BPL luminosity probability density functions, popular within the GRB
literature, as they consistently produce similar quality of fits and as such
neither model is preferred in our conclusions.  

We find that the inclusion of rate, and metallicity density evolution, which are popular
solutions to the issue of underprediction in the GRB formation rate of
high-$z$ bursts, produces marginal improvement in our models however, when
compared to other solutions, are entirely inadequate in explaining the observed
evolution of GRB pulse luminosities. The derived GRB formation 
rate, either incorporating rate density evolution as a Type I-2 model, or as a Type III-1 model,
traces the CSFRD up to high-$z$ and suggests that the parameterisation of
CSFRD models is poor at high redshift, rather than indicating an intrinsic
evolution in the GRB formation rate on top of the CSFRD. We find that within
Type I or Type III models, evolution in the break of the LPDF, as a function
of $(z+1)^{\delta}$, is essential to reproduce the observed $L-z$ GRB pulse
distribution, with $\delta$ exhibiting a strong ($>2$), positive evolution,
consistent with studies that utilise the single brightest GRB pulses. We
evaluated the possibility that this evolution in the break luminosity was down
to the presence of a bimodal population of low/high luminosity GRB pulses,
however our results suggest that a single population of GRBs extending from
the closest, least luminous to the brightest, and furthest GRBs is preferred. We observe that Type III models consistently produce poorer fits to the data than their Type I counterparts, however we conclude that this is an artifact of assuming that components of the Type I progenitor models are known with absolute precision, which is not the case for the CSFRD. To this end we do not attempt to conclude as to the effacy of one method over another.   

We conclude that treating each GRB pulse as an independent event and utilising
the entire GRB pulse population in GRB LF models produces parameters in
excellent agreement to those derived using the single brightest pulse within a
GRB's lightcurve; it is clear that there is no advantage to using solely
the brightest GRB pulse as using all GRB pulses can dramatically improve the
statistics of GRB luminosity functions, and may be extended to investigating
the properties of other intrinsic GRB relationships. Whilst in reality each
pulse cannot be truly independent from another as they are powered from a
single central engine, the relationship between bright, and faint; late, and
early pulses is non-trivial.

\section*{Acknowledgments}
We gratefully acknowledge funding for {\em Swift} at the
University of Leicester by the UK Space Agency. A.A-R's studentship funding is provided by the STFC. We thank the referee for their useful comments and suggestions.

\bibliographystyle{mn2e}
\bibliography{Paper}

\label{lastpage}
\end{document}